\documentclass[
preprint,
superscriptaddress,
amsmath,
amssymb,
showkeys,
aps,
prb,
]
{revtex4-2}

\usepackage{setspace}
\usepackage{amsmath}
\usepackage{graphicx}
\usepackage{hyperref}
\usepackage[mathlines]{lineno}
\usepackage{textcomp}
\usepackage{gensymb}
\usepackage{subfigure}
\usepackage{url}
\usepackage[version=3]{mhchem}
\usepackage{multirow}
\usepackage[american]{babel}
\usepackage{textgreek}
\usepackage[utf8]{inputenc}
\usepackage{floatrow}
\usepackage{amsfonts,amssymb}
\usepackage{miller}
\usepackage{caption}

\begin{document}

\title{High-dose long-time defect evolution in tungsten studied by atomistically informed Object Kinetic Monte Carlo simulations}

\author{Jintong Wu}
  \affiliation{Department of Physics, Post-office box 43, FIN-00014 University of Helsinki, Finland}	
\author{Juan-Pablo Balbuena}
  \affiliation{Department of Physics and Mathematics, Universidad de Alcala, Spain}
\author{Zhiwei Hu}
  \affiliation{CEMHTI, CNRS, UPR3079, University of Orléans, F-45071 Orléans, France}    
\author{Ville Jantunen}
  \affiliation{Department of Physics, Post-office box 43, FIN-00014 University of Helsinki, Finland}  
\author{Marie-France Barthe}
  \affiliation{CEMHTI, CNRS, UPR3079, University of Orléans, F-45071 Orléans, France}   
\author{Maria Jose Caturla}
  \affiliation{Departmento de Fisica Aplicada, Universidad de Alicante, Spain}
\author{Fredric Granberg}
  \email[E-mail: ]{fredric.granberg@helsinki.fi}
  \affiliation{Department of Physics, Post-office box 43, FIN-00014 University of Helsinki, Finland}

\begin{abstract}

Irradiation of materials in nuclear test reactors and power plants is known to alter the properties of the material. The irradiation event happening at pico- or nanosecond time scales are affecting the evolution and properties of the material on macroscopic timescales. Classical Molecular Dynamics simulations, which can capture the cascade event, are typically limited to nanosecond time scales, resulting in high dose rates. To achieve experimental dose rates, larger-scale models like Object Kinetic Monte Carlo are used, while they lack atomistic detail. The exact evolution of cascades in the vicinity of pre-existing defects is known to affect the defects formed, and the structure and morphology of the defects produced are crucial to know for determining macroscopic material behavior. Here we introduce a novel approach to integrate full Molecular Dynamics-based cascades into Object Kinetic Monte Carlo to achieve accurate dose rates, with the atomistic level accuracy of cascade overlap in tungsten. Our study reveals that incorporating full cascades significantly influences defect concentration levels. Not only is the concentration affected, but also the cluster statistics. We observe both that the full cascade can promote vacancy clustering at low temperatures and it can split existing voids at higher temperatures. These effects are missing in conventional Object Kinetic Monte Carlo simulations. This can be especially important in more complex materials, where many cascade-overlap effects are present.

\end{abstract}

\maketitle
\newpage

\section{Introduction}
\label{intro}

The consequences of irradiation in materials, mainly seen as microstructural evolution, originates from the interaction between incident particles and lattice atoms of the material. Radiation damage produced during the collision cascade process leads among others to swelling, embrittlement, and creep~\cite{microstructuralChange_1, microstructuralChange_2, microstructuralChange_3, Mechanic_change}. These mechanical property changing phenomena determine the lifetime of nuclear plant parts. Tungsten (W) has been chosen to play a pivotal role in fusion test reactors under construction, such as the International Thermonuclear Experimental Reactor (ITER)~\cite{ITER1}. Here it will be used as the material facing the plasma inside the reactor, owing to 1) its high atomic mass leading to a low sputtering yield~\cite{Low_sputtering_yield}, 2) its high melting temperature and 3) its high thermal conductivity~\cite{high_temp_W1,high_temp_W2,high_conductivity_W}. 

Irradiation induces significant changes in the microstructure and affects the properties of materials~\cite{microstructuralChange_1, microstructuralChange_2, microstructuralChange_3}. Reza et al.~\cite{W_TGS_exp} used transient grating spectroscopy (TGS) and found that 20 MeV self-ion irradiation showed defect saturation at doses between 0.06 and 0.1 dpa. A positron annihilation spectroscopy (PAS) study by Hollingsworth et al.~\cite{W_PAS_exp_new} showed that vacancies appear as small clusters at low doses at room temperature (RT). Recent research suggests that diffusion of small vacancy clusters, formed by irradiation, can eventually lead to the formation of macroscopic voids, resulting in observable swelling of the material~\cite{crystal_2023}. Voids will also be responsible for material hardening and embrittlement~\cite{wang2016void,dennett2018detecting,hu2016irradiation,hu2016defect,weck2008visualization}. To understand the response of materials to irradiation, a lot of effort has been placed both experimentally and computationally. However, the main drawback of experiments is that, normally, only the end results can be analyzed, and the atomistic features leading to this result are undetectable. To remedy this, simulations with atomistic resolution can be carried out. These simulations will reveal the cascade evolution and the micro-structural changes. Single impact Molecular Dynamics (MD) simulations have revealed the defect production and cluster formation in a single cascade, for instance as a function of energy~\cite{Fik07,Tro11,San19}. Single impact simulations on pre-existing defects have shown that the underlying defect structure dramatically affects the results, and that existing defects can change morphology~\cite{Gra17,byggmastar2019collision,Fellman_2019,gra19}. More recently high-dose MD simulations have been used for understanding the effects of cascade overlap in the nuclear material~\cite{granberg2021JNM,granberg2016mechanism}. Taking advantage of advancements in computational resources, it is now possible to achieve realistic dose regimes, similar to those used in experiments and nuclear reactors. However, the main drawback of MD simulations is the limited timescale that can be achieved, which originates from the need to simulate discrete atoms and their complete trajectories. 

To track the evolution of defects induced by displacement cascades over longer time scales, various Kinetic Monte Carlo (KMC) methods have been developed~\cite{bjorkas2012influence,dalla2005jerk,becquart2006effect,li2021investigating}. Among them, Object Kinetic Monte Carlo (OKMC) is one of the methods able to reach realistic dose rates~\cite{balbuena2019insights,martin2013mmonca}. Regarding tungsten, we know that vacancy defects at RT are practically immobile~\cite{Immobile_Va_1,Immobile_Va_2,Immobile_Va_3}. Hence, in other words, timescale will not effect the vacancy evolution too much. However, at higher temperature the situation is quite different. The migration barrier of vacancies can easily be overcome by thermal activation, which means the voids will start forming and evolving. The microstructure can possibly completely change, when a realistic dose rate is considered at elevated temperatures, compared to what is reachable by MD. 

The OKMC approach is useful for capturing defect concentrations and distributions comparable to experimental conditions, and can directly be compared. In this method, each defect is considered as an object, and its parameterization is crucial, as the energy barrier controls the migration, dissociation, and interaction of defects. Over the decades, numerous studies have shown how the parameterization can affect OKMC simulation results~\cite{castin2017onset,malerba2007object,barashev2000monte,hou2016modification,jansson2013sink}. For instance, Malerba et al.~\cite{malerba2007object}, found that a cubic box shape can cause unrealistic, indefinite migration of self-interstitial atoms, if there is no sink along their path. This is because interstitials in W follow 1D migration, and an indefinite traveling cylinder can lead to a dramatic underestimation of defects as the dose increases. Furthermore, the methods for including defect sinks in the OKMC cell varies. Intrinsic lattice defects, such as dislocation loops, grain boundaries (GB), and impurities, can effectively act as sinks for irradiation induced defects, significantly affecting defect build-up and accumulation. Besides these physically existing sinks, manually adding so-called ``absorbers"~\cite{malerba2007object,ma2024initial} can suppress fast 1D migration in limited boxsizes in OKMC, compensating for model inadequacies~\cite{barashev2000monte}. These numerical sinks can effectively replace physical sinks in tungsten~\cite{ma2024initial}. Although there are multiple ways of adding sinks in OKMC modeling, a comparison of the impact of different sink implementations on defect build-up is rarely studied. 

Currently, the global approach to considering interactions between objects in OKMC is based on the concept of capture radius. A predefined reaction occurs once the distance between two objects falls within this cutoff radius. This method illustrates the overlapping effect in cases of irradiation. However, the true effect of cascades on pre-existing defects over long time scales remains unknown. Studies have shown that on the picosecond (MD) timescale, a cascade can split voids into smaller voids and adjacent dislocation loops~\cite{Fellman_2019}, modify other defect clusters, and that the cascade evolve differently in the vicinity of defects~\cite{Gra17,byggmastar2019collision,gra19}. At realistic time scales, ultimately cascades typically introduce voids, as these structures exhibit lower formation energy compared to planar loops~\cite{neg_vac_vol_3}. The question remains: will the instantaneous effect of cascades on voids last over long time scales and affect the evolution, or will the debris be recaptured, reforming the large voids anyway? Full cascade effects can only be captured by MD simulations, which offer atomistic resolution, a feature absent in the OKMC approach. This suggests that combining atomistically accurate MD cascade simulations with the long-term relaxation of OKMC could achieve atomistic accuracy at experimental dose rates.

In this study, we systematically investigate the effects of absorbers, dose rate, temperature, and defect insertion methods using the OKMC method to understand how these parameters influence the behavior of irradiation defects. Our findings reveal that different methods of setting sinks significantly impact defect accumulation, with the addition of absorbers leading to more realistic defect concentration levels. Based on the determination of realistic parameter settings, we develop a method combining MD and OKMC to evaluate the impact of real cascades on defect buildup. Our results demonstrate that the cascade effect with atomistic accuracy significantly influences defect formation at experimental time scales, particularly in terms of clustering behavior, defect concentration levels, and spatial distribution. When combined with positron annihilation spectroscopy (PAS) and Transmission Electron Microscopy (TEM) results, the improved agreement between our new combination method and experimental studies underscores the necessity of considering the effect of overlapping cascades in high-dose long time scale simulations.

\section{Methods}
\label{sec:method}

\subsection{The object kinetic Monte Carlo approach}

The OKMC simulations were performed using the MMonCa code~\cite{martin2013mmonca}. All defects are considered as objects, such as vacancies, interstitials, and their clusters. Events occur randomly based on predefined probabilities, which can be determined from MD, first-principles calculations and/or experiments. The probabilities of these events are described by the well-known Arrhenius equation~\cite{domain2004simulation}:

\begin{equation}
\nu = \nu_\mathrm{0} \exp \left( -\frac{E_\mathrm{a}}{\kappa_\mathrm{B} T} \right),
\end{equation}

\noindent where $\nu_\mathrm{0}$ is the attempt frequency, $E_\mathrm{a}$ is the activation energy of the event, and $\kappa_\mathrm{B}$ and $T$ are the Boltzmann constant and temperature, respectively. The basic assumptions of the OKMC model used here are essentially identical to those presented in Ref.~\citenum{domain2004simulation}. In MMonCa, it is note-worthy that for the dissociation event, the activation energy is equivalent to the sum of the migration energy of the emitted object and its binding energy with the remaining defect clusters. The binding energy is internally calculated by subtracting the formation energy of the final objects after emission from the formation energy of the initial objects. The range of the interaction between objects depends on their capture radii. More detail of cutoff settings for vacancy-type and interstitial-type objects can be found in Ref.~\citenum{esfandiarpour2022effect}. 

\subsubsection{Grain boundaries and spherical absorbers}
There are plenty of ways setting different sinks in OKMC models, as mentioned in the Introduction section. Studies show a large difference of vacancy concentration at the same dose and temperature conditions~\cite{ma2024initial,li2021investigating}, when different type of sinks are applied. In this study, we compare two approaches: one having only grain boundaries (GBs) and another having GBs with added spherical absorbers, which can be considered as an indirect estimation of sink strength. Migrating defects may interact with virtual GB/absorbers at each migration step with a specific probability, calculated as:

\begin{equation}
P = \frac{k^2 d^2}{2n},
\end{equation}

\noindent where $k^2$ is the effective sink strength and $d=\frac{\sqrt{3}a_\mathrm{0}}{2}$ is the minimum migration distance, and $a_\mathrm{0}$ is the lattice constant at a given temperature. The dimensionality of migration, $n$, is 1 for interstitials and their clusters, and 3 for vacancies and vacancy clusters. For 1D migrating objects, which are, interstitials and the corresponding clusters, dislocation loops, the sink strength for GB ($k_{\mathrm{GB}}^2$) and for absorbers ($k_{\mathrm{ab}}^2$) could be expressed as~\cite{li2021investigating,malerba2007object}

\begin{equation}
k_{\mathrm{GB}}^2 = \frac{15}{{r_{\mathrm{GB}}}^2} , 
\end{equation}
\begin{equation}
k_{\mathrm{ab}}^2 = 6 \left( \pi R^2 N \right) ^2 ,
\end{equation}

\noindent where $r_{\mathrm{GB}}$ is the grain radius~\cite{castin2017onset}, $N$ and $R$ are the density and radius of 1D migrating defects. For a 3D migrating object, $k_{\mathrm{3D}}^2$ can be defined as

\begin{equation}
k_{\mathrm{3D}}^2 = \frac{14.4}{{r_{\mathrm{GB}}}^2}.
\end{equation}

\begin{figure}[]
\begin{center} 
\subfigure{\includegraphics[width=.75\columnwidth]{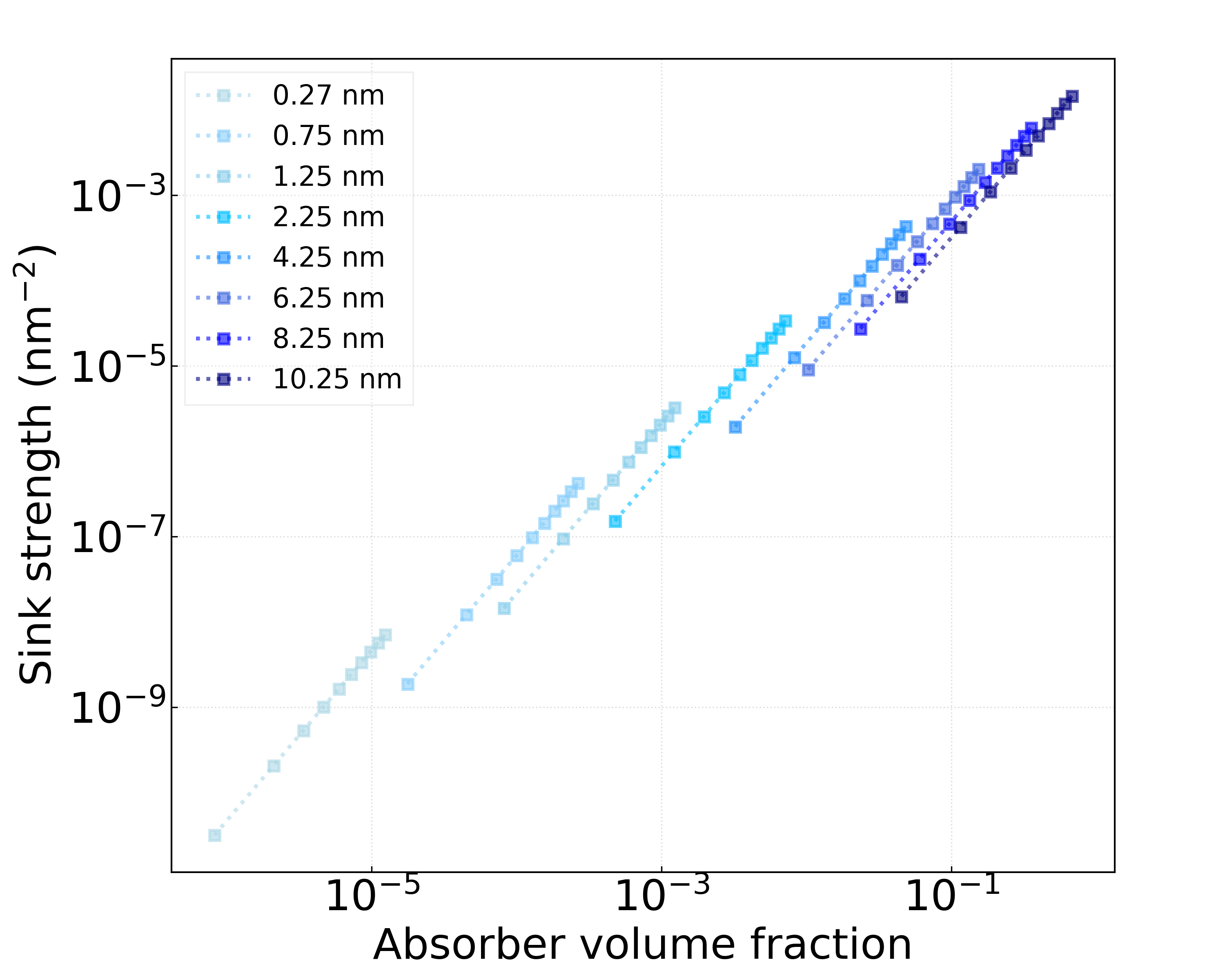}} 
\end{center}
\caption{The sink strength of spherical absorbers for 1D migrating objects as a function of absorber volume fraction.}
\label{sink}
\end{figure}

In this work, we use a grain size radius of 2 $\mu$m, following the settings described in Ref.~\citenum{li2021investigating}. The grain boundary was active for both 1D and 3D migrating defects. However, as we want to suppress the effect of fast 1D migrating interstitials that passes through a small cell many times (this phenomenon is known as the ``size effect"), we applied absorbers to the 1D migrating objects. Experimental positron annihilation results at high temperature show that vacancy clusters are relatively stable~\cite{hu2021effect}. If only virtual GBs are included, the void concentration is underestimated in OKMC~\cite{barashev2000monte, malerba2007object,li2021investigating}, due to recombination with the migrating interstitials, and absorbers can help reduce the recombincation due to the limited cell size. Because computationally expensive MD cascade simulations are performed between each OKMC relaxation in our Combined workflow, the cell used in this study is relatively small. In a smaller cell, the likelihood of highly mobile interstitials repeatedly traveling inside the cell is higher than in a larger cell. Ma et al.~\cite{ma2024initial} investigated the influence of density of absorbers and size of GBs on absorption probability. In our study, we fix the conditions for both parameters, but compare the results with and without the inclusion of absorbers, always considering the GBs. The radius of absorbers varies from 0.27 to 10.25 nm, with densities ranging from $10^{16}$ to $1.5 \times 10^{17}$ cm$^{-3}$, thereby spanning a volume fraction from $1.8 \times 10^{-5}$ to $4.1 \times 10^{-1}$~\cite{malerba2007object, jansson2013sink, hou2016modification, niu2023influence}. Fig.~\ref{sink} shows the cloud of simulation data points grouped by absorbers radii as applied in our simulations.

\subsubsection{Parameters}
\label{sec:OKMCPara}

Due to the 1D diffusion pattern of interstitials in W, as shown in previous studies~\cite{malerba2007object,niu2023influence,ma2024initial,li2021investigating}, the simulation box should have a pseudo-cubic shape~\cite{barashev2000monte}. Here, a non-cubic box with lattice repeats of \(120 \times 130 \times 140 \, a_0^3\) is chosen, with periodic boundary conditions applied in all dimensions. The lattice constant, \(a_0\), was determined by MD simulations at different temperatures. A database of cascade debris by MD simulations is used for inserting point defects with a realistic shape, details found in next section.

\begin{figure}[]
\begin{center} 
\subfigure{\includegraphics[width=.75\columnwidth]{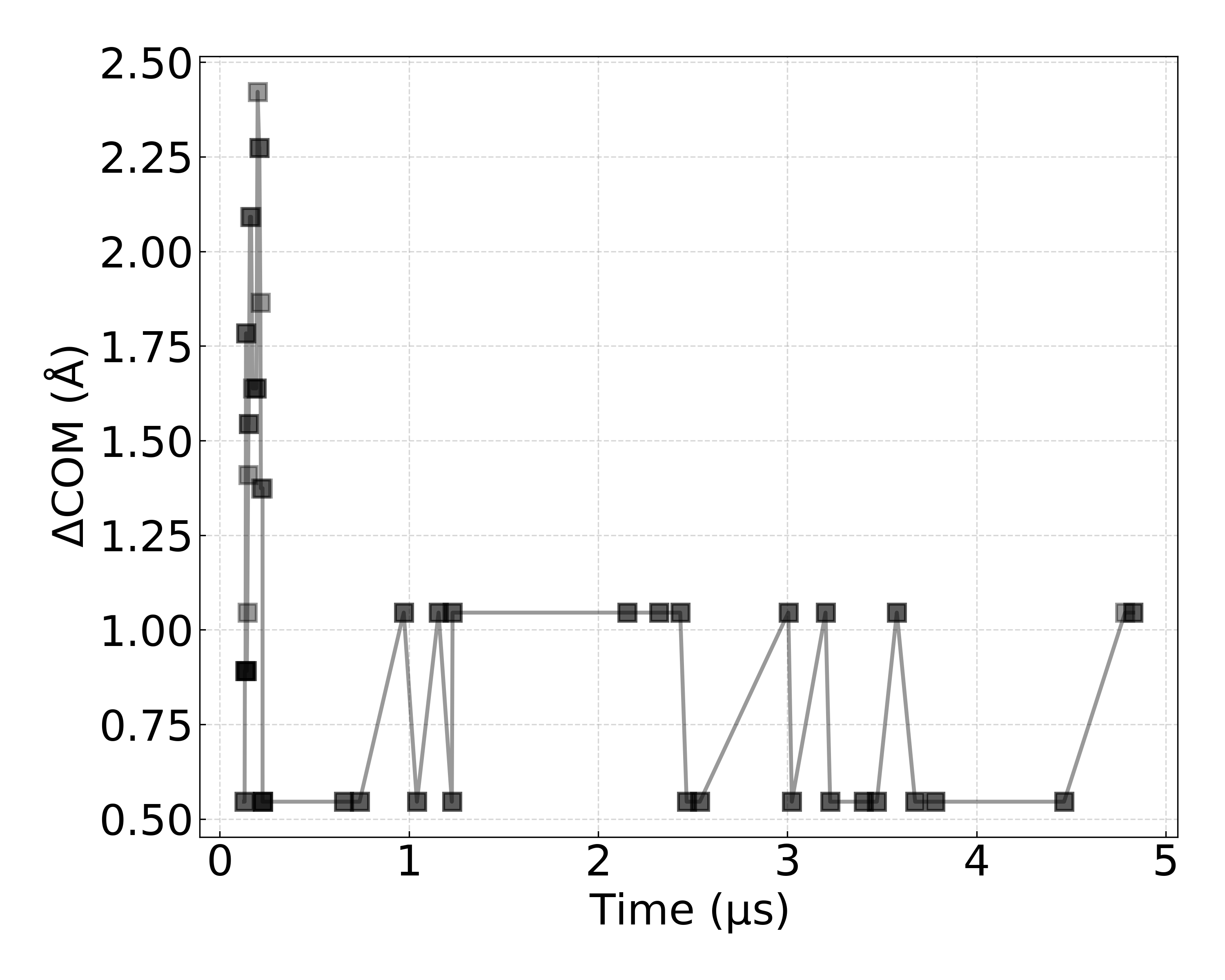}} 
\end{center}
\caption{Relative distance of a V5-cluster to its original position as a function of time at 1500K.}
\label{fig:kart}
\end{figure}

Two main types of defects are considered in our OKMC model: interstitials and vacancies. Leveraging previous work~\cite{li2021investigating}, the migration energies of V1 to V4 are determined, with a migration attempt frequency of $6 \times 10^{12}/$s. Large vacancy clusters (size $>$ 4) are considered immobile. The authors of Ref.~\citenum{li2021investigating} considered vacancy clusters larger than V4 to be immobile. We tested this assumption using an off-lattice KMC code kART~\cite{kart}. We simulated an otherwise perfectly crystalline system with a single V5 cluster at 1500K. We did not observe any noticeable migration of the defect during the 5-microsecond simulation time. The slight oscillation of the cluster center, shown in Fig.~\ref{fig:kart}, is due to individual vacancy occasionally jumping to the nearest neighbor site or, in rare cases, a few sites away before returning to the cluster.

The interstitial and its clusters can efficiently migrate along one of the four \hkl<111> directions with a migration barrier of 0.013 eV~\cite{niu2023influence}, which is much lower than that of vacancies. The rotation energy for a mono-interstitial traveling between different \hkl<111> paths is 0.38 eV~\cite{derlet2007multiscale}, while interstitial clusters are not allowed to rotate due to their high rotation energy~\cite{zhou2014transport}. Additionally, the formation energy of interstitials and vacancies of different sizes is taken from Ref.~\citenum{neg_vac_vol_3}. 

In the present work, three different variables are tested and compared to each other:

\begin{enumerate}
    \item Spherical absorbers with Virtual GBs vs. Only Virtual GBs: We compare the effects of including spherical absorbers along with virtual GBs against using only virtual GBs to clarify the impact of absorbers on the defect accumulation process.

    \item Cascade-induced defect shape insertion: We investigate the influence of inserting defects with cascade shape (CS), which are the result of primary knock-on atom (PKA) MD simulations with an energy of 30 keV, allowing direct comparison with a previous study~\cite{wu2023atomistic}. In this scenario, defects are distributed along the path of the PKA atom. This approach contrasts with the Frenkel Pair (FP) insertion method, recently proven to be efficient for speeding up simulations on MD timescales~\cite{derletCRA,MasonPRL}. In this method random FPs are inserted without any restriction on defect positions. We examine whether CS affects defect evolution on realistic time scales.

    \item Different dose rates: According to Ref.~\citenum{hu2021effect}, the dose rate in ion irradiation is around \(3 \times 10^{-5}\) dpa/s (with a displacement energy threshold of 90 eV). Using the classical Norgett-Robinson-Torrens displacements per atom (NRT-dpa) model~\cite{NRT_N,NRT_R,NRT_T}, a 1-second relaxation between adjacent cascade events results in a similar dose rate of about \(2.3 \times 10^{-5}\) dpa/s for our system size. If the duration between two cascade events is extended to 1 minute, the corresponding dose rate becomes \(3.8 \times 10^{-7}\) dpa/s, which is comparable to neutron irradiation conditions~\cite{ma2024initial,li2021investigating,castin2017onset}. We compare how different dose rates affect the results and clustering of defects.
\end{enumerate}

For all parameter sets, three independent runs were carried out for statistics.

\subsection{Irradiation simulations in MD}

The cascade defect database used in OKMC simulations were conducted with LAMMPS~\cite{lammps}, implemented with an adaptive timestep~\cite{adaptive_timestep1,WS,adaptive_timestep3}. The dimensions of the cell were $120 \times 120 \times120$ conventional BCC unit cells, resulting in 3.456 million atoms. Periodic boundary conditions were applied in all dimensions. A friction force was applied to all atoms according to the energy loss table provided by ``Stopping and Range of Ions in Matter" (SRIM)~\cite{srim} code to include effects of electronic stopping power, when the kinetic energy of an atom exceeds 10 eV~\cite{elstop}. We used the embedded atom method (EAM) potential by Mason et al.~\cite{Mason_2023} for all MD simulations. A 30 keV PKA was introduced at the centre of the system, in a random direction. Most of the cell was able to evolve without any constraints, except a two nanometer thick layer at the borders, which were thermally controlled to a certain temperature. No pressure control was active during this simulation. Each cascade was followed for 30 ps, followed by a relaxation period of 10 ps with temperature control at desired temperature on all atoms and a pressure control to zero overall pressure.

For the database used in OKMC, 100 random PKAs were simulated and the final defects were obtained by the Wigner-Seitz (W-S) method~\cite{WS}, at each temperature. To obtain the defect evolution according to pure MD, the same methodology was used, however, instead of single PKAs impacting a pristine cell, the cells were subjected to 4000 PKAs, to achieve the dose of $\sim$0.1 dpa (with a threshold displacement energy of 90 eV) according to the classical NRT-dpa model~\cite{NRT_N,NRT_R,NRT_T}. Between each PKA, the cell was randomly shifted to obtain a homogeneous irradiation, according to previous MD high-dose studies~\cite{granberg2016mechanism,granberg2021JNM}. In this case the debris from the last impact was always present when staring the next PKA. Three independent pure MD runs were carried out at all investigated temperatures for statistics. In the combined MD and OKMC workflow, described in the next section, the MD cascade part is identical to the one for the pure MD simulation.

\subsection{Combination of atomistic MD cascades with the long-term relaxation of OKMC}

\begin{figure}[]
\begin{center}
\subfigure{\includegraphics[width=.8\columnwidth]{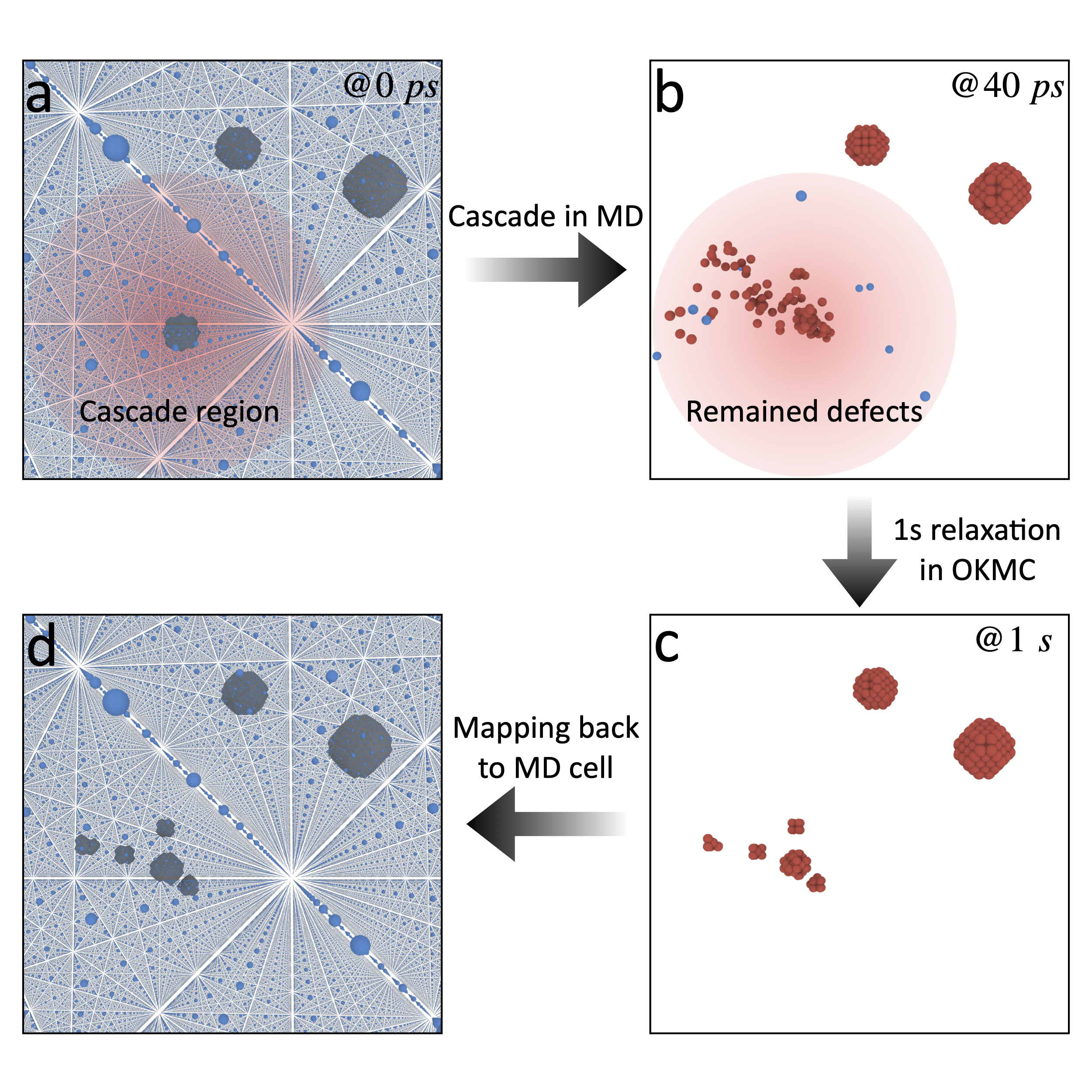}} 
\end{center}
\caption{Schematic representation of the combination workflow. Defect meshes representing voids are shown in dark grey, rendered using OVITO. Blue atoms indicate W atoms with a BCC atomic arrangement. (a) Initial cascade event; (b) Defect distribution at 40 ps post-irradiation; (c) Defect distribution after a 1-second OKMC relaxation, with red and blue atoms denoting vacancies and interstitials detected by the W-S method; (d) Mapping back to the atomistic configuration in preparation for the next cascade event.}
\label{CombSketch}
\end{figure}

Fig.~\ref{CombSketch} illustrates the concept of our combination method with atomistic full MD cascades with the long term evolution of OKMC. Previously, OKMC simulations are applied by using a databased of collision cascades where the damage of a single PKA in a pristine sample obtained from MD simulations~\cite{malerba2007object,ma2024initial,li2021investigating}. Here, atomistically accurate MD cascades are used on cells with all pre-existing defects, instead of inserting single PKA from a database. The cascade events are performed in an MD cell, as shown in Fig.~\ref{CombSketch}(a). After the collision cascade, defects evolve on the picosecond timescale to reach a temporary equilibrium state, Fig.~\ref{CombSketch}(b), reached by full MD.

Next, the defects are identified by using the W-S method and inserted into an OKMC cell, which is evolved a similar amount of relaxation as experimental samples, corresponding to the comparable dose rate, as shown in Fig.~\ref{CombSketch}(c). Finally, the remaining defects (objects) are mapped back to the atomistic cell (with full atomistic resolution), followed by a 10 ps relaxation at the correct lattice constant at a certain temperature, preparing for the next round of overlapping cascades, Fig.~\ref{CombSketch}(d). This iteration continues until the desired dpa level is achieved, allowing for comparison with traditional OKMC results. In all cases, three different temperatures (RT, 500\degree C, and 700\degree C) are studied, with three independent runs implemented for each set to obtain statistics.

The ``Open Visualization Tool" (OVITO)~\cite{ovito} was used for visualization. For mapping back and forth between the pure OKMC and the MD cells, the defects were detected based on W-S analysis~\cite{WS}. Cluster size distribution analysis was performed on the obtained point defects by W-S. In cases where a cluster contains both vacancies and interstitials, we determined the net content of defects to accurately assess the defect size. The cutoff values for vacancies and interstitials are the intermediate ones between the second nearest neighbor (2NN) and 3NN, and between the 3NN and 4NN, respectively~\cite{Vac_cutoff,Inter_cutoff}.

\subsection{Positron annihilation spectroscopy and simulated annealing algorithm}
\label{PAS}

PAS is commonly utilized to detect vacancy-type defects in materials~\cite{zhiwei1,zhiwei2,zhiwei3}. It is highly sensitive, capable of detecting from single vacancies (V1) to vacancy clusters (VN), and usually serves as a complementary technique for characterizing defects that are not visible by a transmission electron microscope~\cite{hu2021effect,zhiwei5}. Once inside the material, positrons undergo thermalization and diffusion within the matrix.

During diffusion, positrons favor being trapped at vacancy defects due to the repulsive force from the nuclei, then annihilating with an electron in the vicinity of the vacancy, i.e. localized state. If not trapped, the positron can annihilate with electrons in the the crystal lattice i.e. delocalized state. The positron-electron annihilation emits in most cases two gamma-rays with an energy spectrum (or momentum spectrum) broadened at 511 keV (or $m_\mathrm{e} c$, $m_\mathrm{e}$: mass of electron, and $c$: velocity of light). The annihilation characteristics, or line-shape parameters \emph{S} (fraction of the low momentum events) and \emph{W} (fraction of the high momentum events) characterize positron annihilation states. Hu et al. employed PAS and TEM to characterize self-ion irradiated W at different temperatures, the \emph{S} and \emph{W} parameters differ for different sample purities, revealing that light-element impurities impeded the clustering of vacancies~\cite{hu2021effect}. We, therefore, used the high purity samples results in~\cite{hu2021effect,hu2024revealing} to compare with our simulation results. Recently, the combination of PAS experiments and first-principles calculations has shown encouraging results in determining the annihilation characteristics (\emph{S}-\emph{W} parameters) of $V_1$ to $V_N$~\cite{zhiwei5}. 

Combining a commonly-used global optimization approach, in our case the simulated annealing (SA) algorithm~\cite{zhiwei8}, with the positron trapping model~\cite{zhiwei6,zhiwei7}, the most probable annihilation fraction, $f_i$, of different types of $V_i$ can be obtained~\cite{letter}. Tens thousands of iterations are conducted to search the most likely approximate global distribution of $f_i$. Then, according to the trapping model, the relation between the $f_i$ and the trapping rate $k_i$ of each type of vacancy was expressed by

\begin{equation}
f_i = \frac{k_i}{\lambda_\mathrm{L} + \sum_{i=1}^{n} k_i}.
\end{equation}

As estimated in Ref.~\citenum{hu2021effect}, the annihilation fraction in the delocalized states is about 1 \% or less, hence the annihilation rate at the lattice i.e. $\lambda_\mathrm{L}$ is negligible before the $k_\mathrm{i}$. Thus, the annihilation fraction is simplified to the ratio $\frac{k_i}{k_{\text{tot}}}$. The total trapping rate $k_{\text{tot}}$ of positrons at vacancies was estimated from the effective positron diffusion length~\cite{W_PAS_exp_new} using the VEPFIT program~\cite{zhiwei10}. Once the $k_i$ is available, the concentration of each size of vacancy cluster i.e. $c_i$ can be calculated as

\begin{equation}
k_i = \mu_i c_i,
\end{equation}

\noindent where $\mu_i$ is the specific trapping coefficient at each type of vacancy cluster, which can be found in~\citenum{letter}. By summing up the concentration of different sizes of voids, we obtain the estimation of total concentration of vacancies at different temperatures by PAS.

\section{Results}

In the Results section we will focus on the vacancy formation and clustering under different conditions. In all OKMC and Combination simulations the interstitials did either get combined with vacancies or got absorbed by the GBs or the Absorbers. This is due to their very high mobility even at low temperatures. Therefore, they are not discussed in the following sections, except in the context of why vacancy evolution is as it is. In the full MD cases, due to the periodic cell and no sinks, the interstitials are present. They mainly form large 1/2\hkl<111> loops as the dose increases. For detailed analysis of interstitials in the full MD case, please see previous studies, such as Refs.~\citenum{granberg2021JNM} and~\citenum{Granberg2022}.

\subsection{Defect formation in W under different OKMC parameterization}
\subsubsection{Influence of cascade shape}

\begin{figure}[]
\begin{center} 
\subfigure{\includegraphics[width=.9\columnwidth]{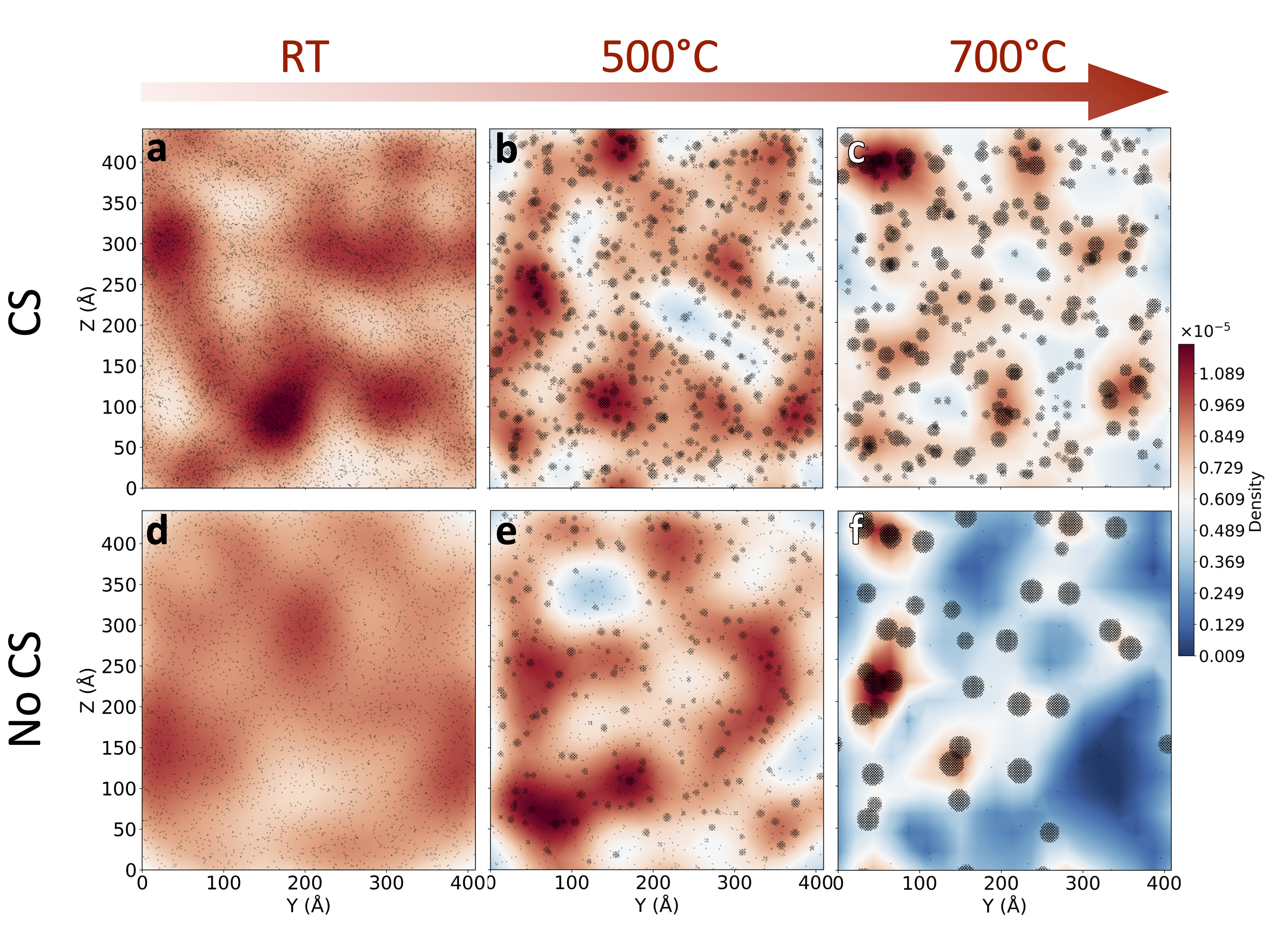}} 
\end{center}
\caption{2D density maps illustrating vacancy distribution at 0.1 dpa with and without CS enabled at three different temperatures. Black dots indicate the projection of vacancies aligned with the \hkl[100] direction.}
\label{HeatMap}
\end{figure}

To investigate the influence of CS at different temperatures, the other conditions are kept constant with a 1-second adjacent cascade duration and with absorbers enabled. To show a straightforward comparison, the vacancy clusters at 0.1 dpa are shown in Fig.~\ref{HeatMap}, where the density is color-mapped in a 2D projection.

At RT, without CS, the voids exhibit a relatively homogeneous distribution. However, as seen in Fig.~\ref{HeatMap}(a), even though rigorous clustering analysis does not reveal large clusters, the CS case shows a denser local distribution of vacancies. As temperature increases, it becomes evident that at 500\degree C, having a shape to the inserted debris produces larger voids compared to the one without CS, as shown in Fig.~\ref{HeatMap}(b) and (e). Interestingly, at 700\degree C, the trend reverses: the non-CS case results in a much larger void distribution than the CS-insertion case, as illustrated in Fig.~\ref{HeatMap}(c) and (f).

On realistic timescales, a 200\degree C temperature difference significantly impacts defect clustering. At 500\degree C, vacancies start becoming mobile, though the probability of motion remains relatively low. Clustering tends to occur but is much less pronounced compared to 700\degree C. Comparing Fig.~\ref{HeatMap}(b) and (e), the distribution of vacancies and their clusters is quite similar. Thus, the CS consideration leads to a lower chance of recombination, resulting in larger local voids compared to random defect insertion in the cell.

At 700\degree C, the mobility of vacancies is greatly enhanced compared to the case at 500\degree C. Although large voids are immobile, dissociation causes them to emit mono-vacancies, allowing large clusters to jitter around as they grow. When vacancies are inserted randomly, the chance of local clustering is much lower than in the CS case. A previous study~\cite{byggmastar2019collision} indicates that the local molten region where FPs are primarily generated by a single cascade is roughly a spherical region with a radius of about 30 Å in W at 30 keV case. The denser point defect insertion in the CS case explains the formation of smaller but greater number of local voids, as seen in Fig.~\ref{HeatMap}(c). The FP insertion method allows vacancies to travel throughout the cell rather than being captured due to close insertion distances. This is supported by migrating logs from OKMC simulations, showing that, on average, for every 1-second relaxation duration, migrating events in the random insertion case (no CS) occur about 50 times more frequently than in the CS case.

\subsubsection{Influence of dose rate}

\begin{figure}[]
\begin{center} 
\subfigure{\includegraphics[width=.9\columnwidth]{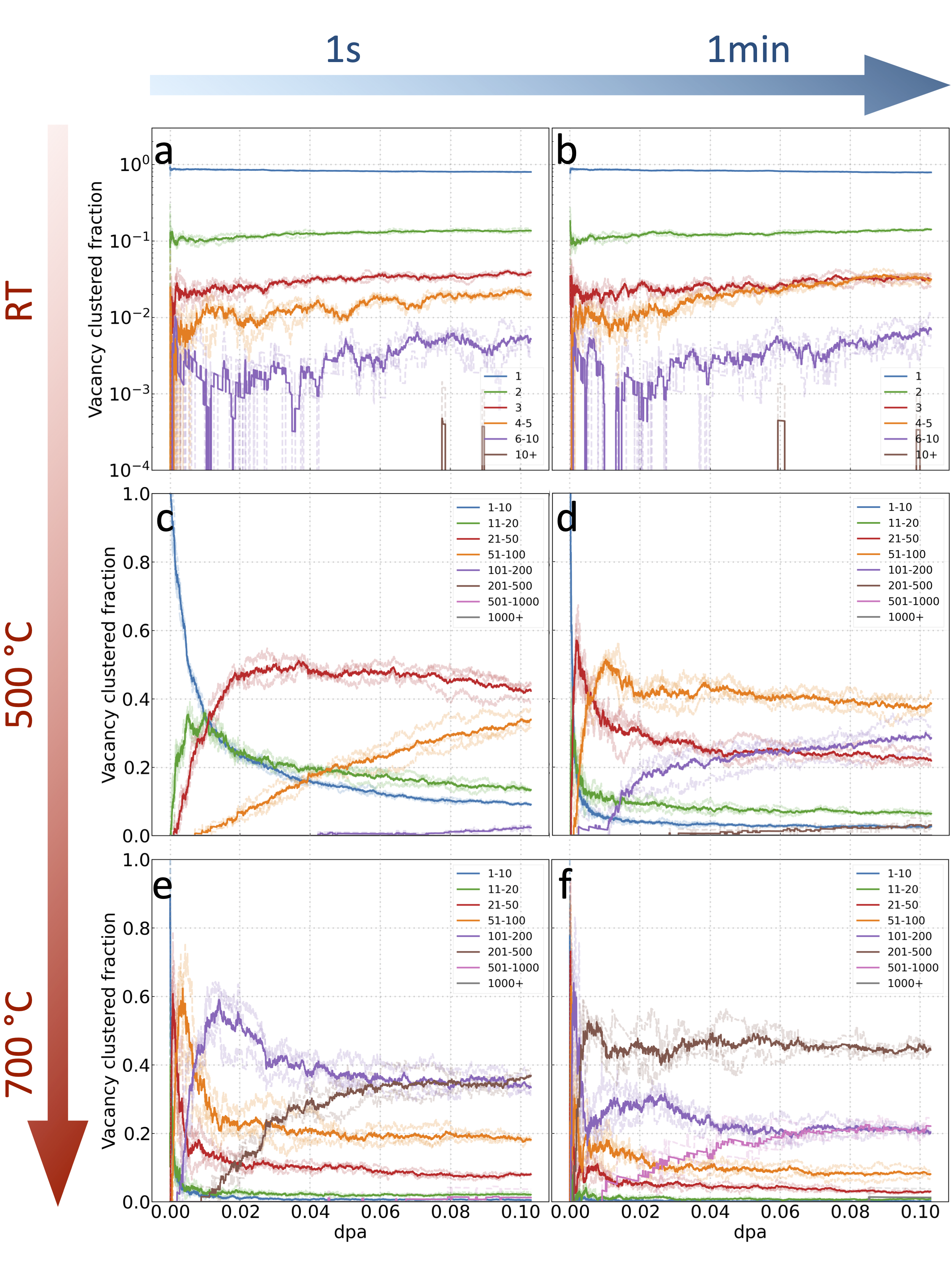}} 
\end{center}
\caption{Fraction of vacancies found in different-sized vacancy clusters. The different coloured lines represent different cluster sizes and the y-axis the average number of vacancy atoms in the respected cluster size range. Dashed lines represent three independent runs and the solid line shows the average value.}
\label{VacFrac}
\end{figure}

From the previous section, we know that cascade shape significantly affects the spatial distribution of clusters, and that it is more realistic than random FP insertion at these high energies. We choose conditions where both CS insertion and absorbers are enabled to investigate the influence of dose rate. Fig.~\ref{VacFrac} shows the clustered vacancy fraction as a function of dose under different conditions. Note that logarithmic scaling is applied to the RT case, as seen in Fig.~\ref{VacFrac}(a) and (b). A very similar trend in cluster fraction evolution is observed at RT. This is due to the high migration barrier of vacancies, making them stable and rendering the dose rate effect negligible at RT.

The influence of dose rate becomes significant at higher temperatures. At 500\degree C, clusters ranging in size from 1 to 10 show a much quicker drop during a 1-minute relaxation (corresponding to a $10^{-7}$ dpa/s dose rate) compared to a 1-second relaxation (corresponding to a $10^{-5}$ dpa/s dose rate), as shown in Fig.~\ref{VacFrac}(c) and (d). Meanwhile, the larger clusters (in the size range of 21 to 50) obviously increase more in the 1-minute case compared to the 1-second case. This indicates that a lower dose rate results in quicker cluster growth as a function of dose. For clusters sized 21 to 50, the peak fraction of 0.6 is reached at about 0.005 dpa, followed by a rapid decrease, converging at around 0.2, as shown by the red curve in Fig.~\ref{VacFrac}(d). Meanwhile, larger clusters (101-200) continue to grow and have not saturated by the time the dose reaches 0.1 dpa. In contrast, at 500\degree C with the 1-second duration, cluster growth is relatively slower, with smaller clusters predominating. The fraction of large clusters (101-200) is almost nonexistent, whereas in the 1-minute case, it exceeds 0.2 at 0.1 dpa. The predominant cluster size in the 1-second case is 21-50, decreasing even at the end of the investigated dose range, while clusters sized 51-100 exhibit rapid growth, surpassing the growth rate of the 1-minute case at 0.1 dpa. This suggests that clustering differences between dose rates might diminish as the dose further increases.

At 700\degree C, the influence of dose rate is also apparent. In the 1-second case, clusters in the 101-200 size range rapidly predominate at around 0.01 dpa, then transition smoothly to larger clusters (ranging 201-500) until they reach almost the same fraction level at the end. With a 1-minute adjacent relaxation, vacancies cluster more rapidly, and clusters of all sizes show a converging trend once the dose reaches about 0.06 dpa. The cluster size between 201-500 immediately becomes the most populated at the very beginning, around 0.005 dpa, and remains so until the end. Larger clusters (501-1000) also appear but do not form throughout the process in the 1-second case.

\subsubsection{Influence of spherical absorbers}

\begin{figure}[]
\begin{center} 
\subfigure{\includegraphics[width=.7\columnwidth]{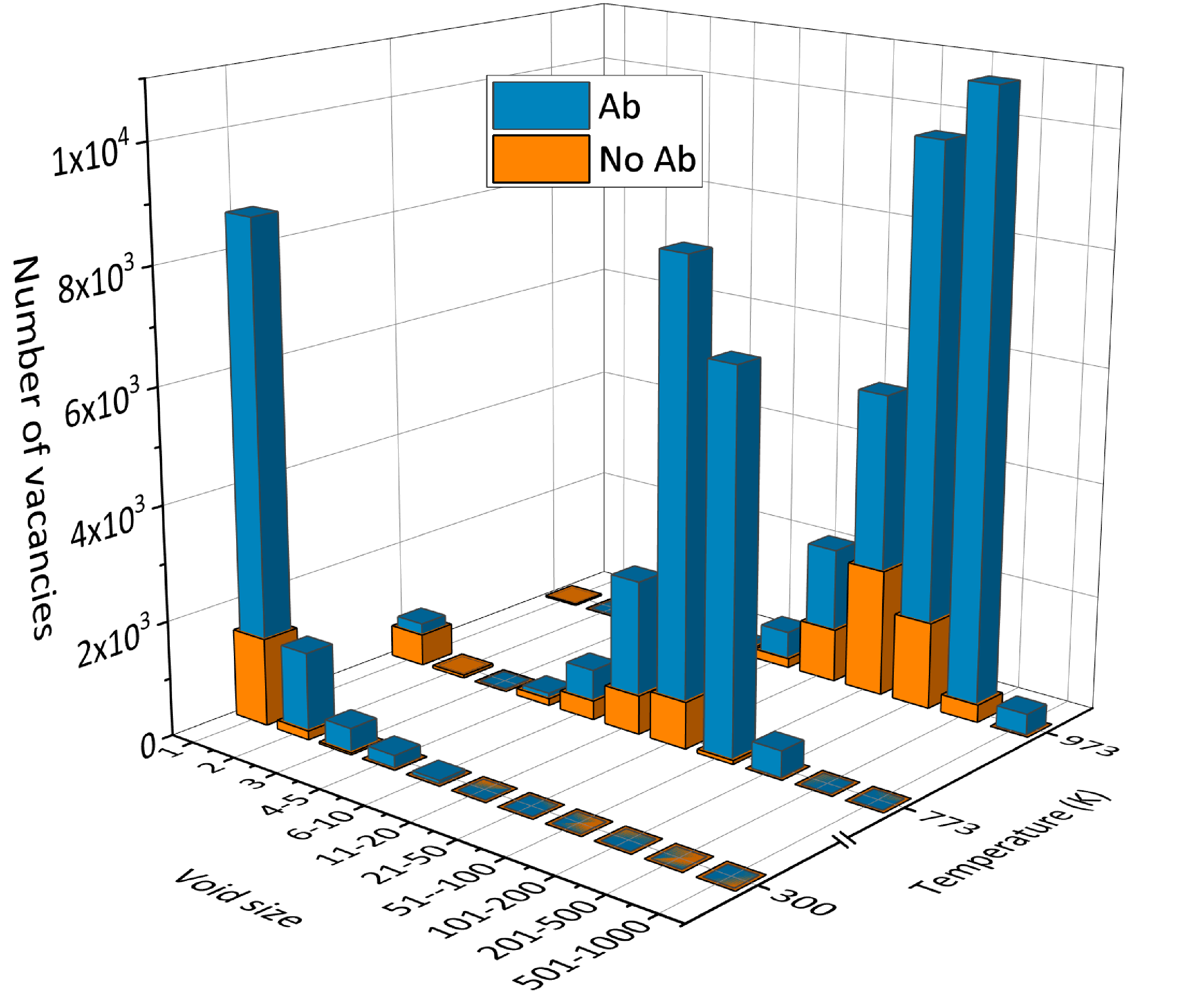}} 
\end{center}
\caption{Size distribution of vacancy clusters with and without spherical absorbers enabled at different temperatures.}
\label{AborNot}
\end{figure}

Fig.~\ref{AborNot} shows the size distribution of vacancy clusters in irradiated W up to 0.1 dpa at different temperatures. The conditions are fixed with CS insertion and a 1-second adjacent cascade duration to assess the influence of absorbers at various temperatures. At RT, clusters do not exceed a size of 10 in either case. The absorbers-enabled set shows significantly much more clusters than the one without absorbers, with mono-vacancies being the predominant form. This is due to the higher number of surviving interstitials in the absorber-disabled case, leading to increased recombination.

As temperature increases with absorbers enabled, the peak of the histogram shifts from the 21-50 size range at 500\degree C to the 201-500 size range at 700\degree C. In the absorbers-disabled case, the peak size range slightly differs, with the largest proportion of clusters being 51-100 at 700\degree C, for instance. The common thing is the trend of forming the larger cluster with the increasing of temperatures is not affected by enabling the absorbers or not. While absorbers do not significantly alter the predominant cluster size distribution, they do affect the absolute number (concentration) of surviving clusters. The strong sink strength provided by absorbers effectively addresses the size effect issue in OKMC modeling in the W case: the virtual absorbers acts as a trap, capturing and removing rather mobile interstitials from the periodic cell, thereby avoiding underestimation of vacancy concentration level. However, it has to be noted that this may not be a universal solution for other materials. For example, in iron, immobile C15 interstitial-type clusters~\cite{marinica2012irradiation} can significantly alter microstructural evolution even without considering additional sinks, like impurities or absorbers, as C15 itself can serve as a natural sink to trap and recombine with mobile vacancies. Therefore, this remains an open question requiring a further study.

Fig.~\ref{Overall} presents the evolution of vacancy concentration across various cases. Here, to assess the impact of absorbers, we focus on Figs.~\ref{Overall}(a), (c), and (e), which represent the pure OKMC scenarios. The significant differences in accumulated vacancy concentration across all three temperature cases reveals the strong impact of enabling absorbers. At high temperatures, even though local voids form in contrast to the monovacancy dominated cases, the absence of absorbers still results in a lower absolute vacancy concentration, showing trends of rapid convergence, as seen in Figs.~\ref{Overall}(c) and (e). This indicates a relatively lower likelihood of recombination with interstitials compared to cases dominated by monovacancies. When compared with experimental data, it is clear that disabling the absorbers greatly influences the vacancy concentration in all cases. Therefore, we argue that, at least for W irradiation simulations, activating the absorbers is essential for achieving realistic vacancy concentration levels. More details about the experimental results are found in the following sections.

\subsection{Results by the atomistically informed OKMC model}
\subsubsection{Vacancy concentration evolution}

\begin{figure}[]
\begin{center}
\subfigure{\includegraphics[width=.9\columnwidth]{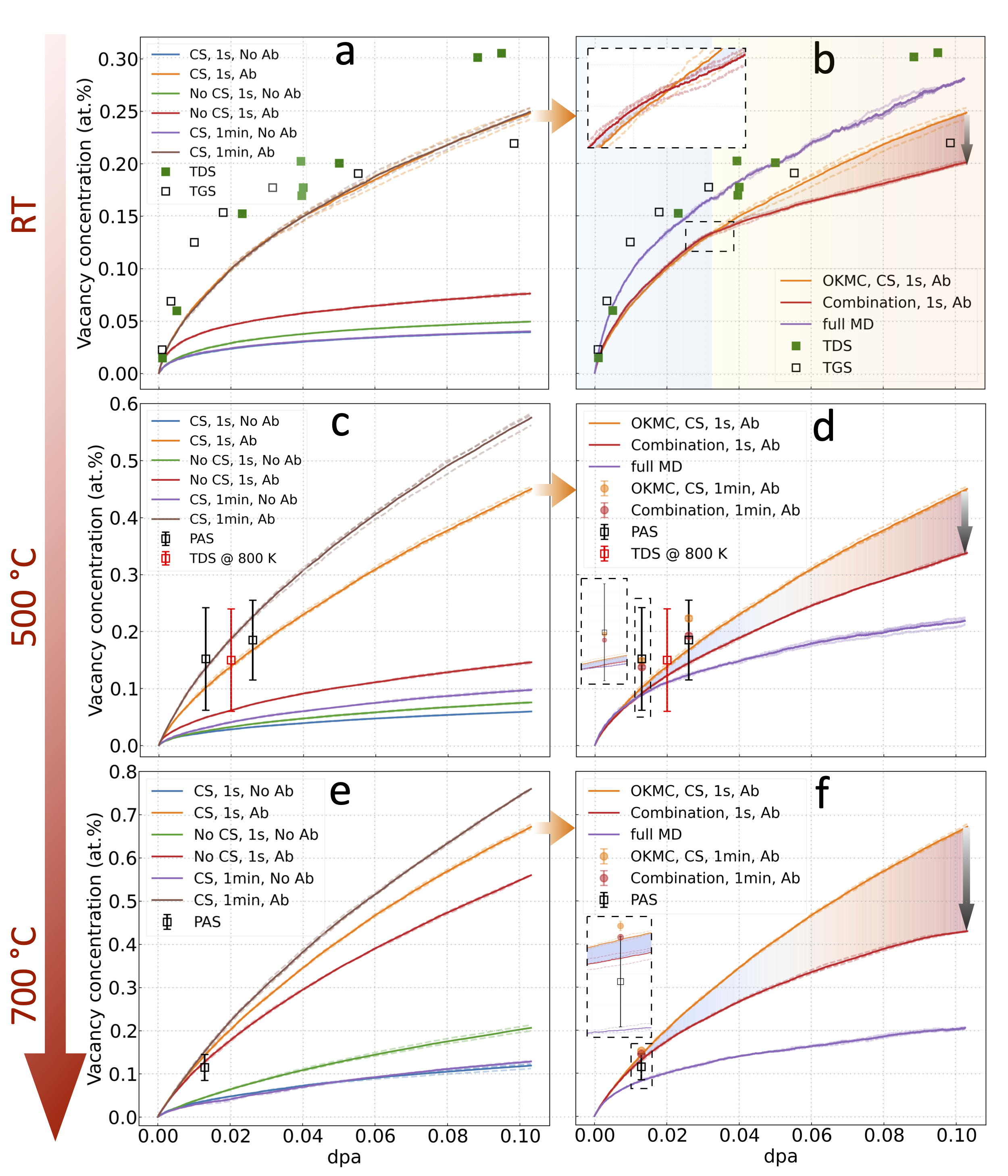}} 
\end{center}
\caption{Evolution of total vacancy concentration as a function of dose under various conditions. Left panel: pure OKMC simulations with different parameterizations. Right panel: comparison of the pure OKMC with the combination method, using full MD data and experiments as a reference. The gradient color fill shows the concentration differences between the original OKMC and the combination method. For each set, dashed lines represent three independent runs, while the solid line indicates the average value. The PAS raw data points (\emph{S}-\emph{W} parameters) are obtained from Ref.~\citenum{hu2021effect}, and the vacancy concentration estimated in this study by the SA method~\cite{letter}. The TDS and TGS experimental results at RT are obtained from Refs.~\citenum{MasonPRM_VoidDetection,wang2023dynamic,simmonds2019isolating,pevcovnik2020effect,wielunska2020deuterium,schwarz2018influence} and Ref.~\citenum{W_TGS_exp}, respectively. The TDS data at 800 K is obtained from Ref.~\citenum{markelj2024first}, here we assume each vacancy could trap either 1 or 5 D atoms, the average value is taken to show the assumed vacancy concentration (therefore the huge errorbar).}
\label{Overall}
\end{figure}

At RT, the vacancy concentration under various conditions is compared to experimental data obtained through the thermal desorption spectrometry (TDS) of hydrogen(assuming one vacancy can capture five deuterium atoms on average)~\cite{MasonPRM_VoidDetection,wang2023dynamic,simmonds2019isolating,pevcovnik2020effect,wielunska2020deuterium,schwarz2018influence} and transient grating spectroscopy (TGS)~\cite{W_TGS_exp}, as shown in Fig.~\ref{Overall}(a) and (b). Two sets of conditions from the pure OKMC simulations yield results that align with experimental estimates: those with CS-insertions and absorbers, whether the adjacent cascade duration is 1 second or 1 minute. This similarity is primarily due to the high migration barrier~\cite{li2021investigating}, which renders vacancies relatively immobile at RT. Thus, the concentration remains consistent despite changes in dose rate when other conditions are identical. This is also demonstrated by the overlapping curves of the CS-1s-no-absorbers (blue) and CS-1min-no-absorbers (purple) cases in Fig.~\ref{Overall}(a). Based on the correct ballpark parameter settings, we select the set with an experimentally compatible dose rate (1-second cascade duration) with absorbers for comprehensive combination runs, facilitating a thorough comparison at all dpa levels with the pure OKMC cases, as shown in the right panel of Fig.~\ref{Overall}, where the full MD results are also plotted as reference. Fig.~\ref{Overall}(b) shows the RT case of this direct comparison. An eye-catching phenomenon is observed: at RT, a sharp change in the slope of the vacancy concentration curve occurs only with the combination method, around 0.03 dpa, as shown in the zoomed-in panel. This distinct accumulation slope change is also observed at approximately the same dpa level in both the TGS and TDS experiments. Two different background colors are used to distinguish the state before the distinct slope change and after.

The same pure OKMC set was chosen for a comprehensive investigation using a combination method to facilitate a thorough comparison at high temperatures. Specifically, this was done with a 1-second cascade duration with absorbers enabled. Our results indicate that only at RT will there be a crossover between the pure OKMC and the combination method. At high temperatures, the combination method shows a lower concentration level from the beginning of irradiation, as shown in Fig.~\ref{Overall}(d) and (f). Most OKMC cases without CS insertion and those without absorbers significantly underestimate the vacancy concentration level, as seen in Fig.~\ref{Overall}(c) and (e). Specifically, the PAS results for the 0.013 and 0.025 dpa cases align better with the 1-minute and 1-second simulation cases at 500\degree C, respectively, as illustrated in Fig.~\ref{Overall}(d). From the 500\degree C to the 700\degree C case, the difference between the pure OKMC and the combination method becomes more pronounced with the increasing of damage levels. In all three cases, the MD results show nearly identical accumulation levels, while an increase in vacancy concentration in both the pure OKMC and combination methods is observed. 

\subsubsection{Cascade induced clustering behavior}

\begin{figure}[]
\begin{center} 
\subfigure{\includegraphics[width=.9\columnwidth]{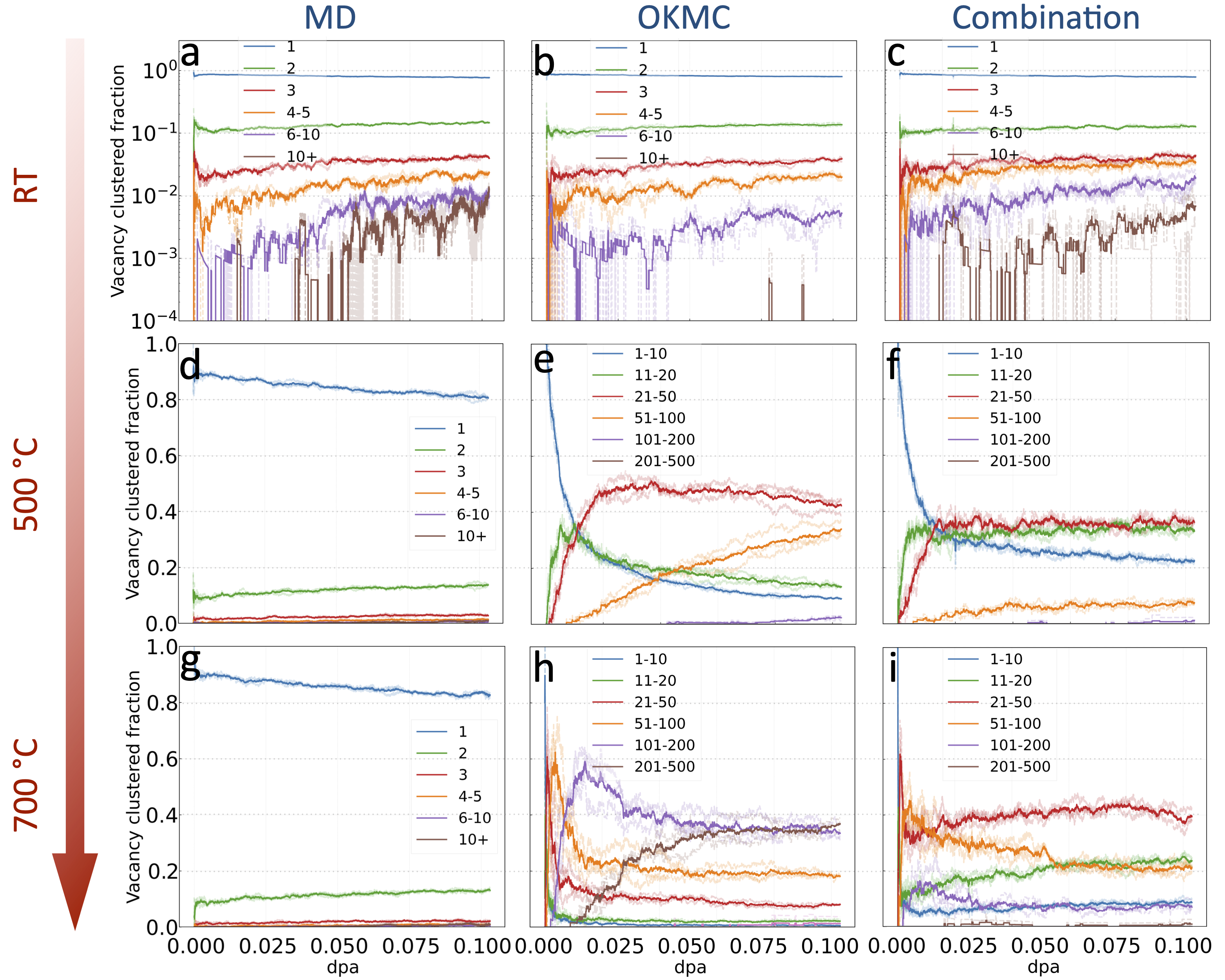}} 
\end{center}
\caption{Vacancy clustered fraction as a function of dose at different temperatures. The left panel: the full MD case; the middle panel: the OKMC case; the right panel: the combination case. The different colored lines indicate various cluster sizes, with the y-axis representing the average number of vacancies within each respective cluster size range. Dashed lines depict results from three independent runs, while the solid line shows the average value.}
\label{Comb_clus}
\end{figure}

To better understand the differences between the traditional OKMC approach and our combination method regarding void evolution and accumulation, we quantified the clustering behavior at different doses. Fig.~\ref{Comb_clus} illustrates the distinct clustering behavior of the three methods, i.e. full MD as a benchmark shown on the left side, pure OKMC shown in the middle and our combination method shown on the right.

At RT, the full MD results show stable-growing clusters larger than 10 appearing after a dose of about 0.03 dpa (Fig.~\ref{Comb_clus}(a)). In contrast, this trend is absent even at the highest dose (0.1 dpa) investigated in this work in the pure OKMC case (Fig.~\ref{Comb_clus}(b)), suggesting that the real cascade effect is crucial for forming larger clusters at RT in W. Interestingly, large clusters (greater than 10) appear earlier, around 0.013 dpa, in the combination method (Fig.~\ref{Comb_clus}(c)), indicating that full cascade effect combined with long-term relaxation promote larger cluster formation. 

At 500\degree C, significant differences among the three approaches are evident. Due to the lack of a realistic timescale, the full MD results remain similar to those at RT (Fig.~\ref{Comb_clus}(d)). This due to the vacancies not being mobile on MD timescales. Both pure OKMC and our combination methods predominantly show cluster size ranges from 21 to 50. However, rapid growth of clusters larger than 50 is observed only in the pure OKMC case (Fig.~\ref{Comb_clus}(e)) even at the highest dose (0.1 dpa), whereas the combination method shows a stable fraction distribution of various cluster sizes after reaching 0.025 dpa (Fig.~\ref{Comb_clus}(f)). 

At 700\degree C, the most significant cluster size in the pure OKMC case shifts from 21-50 at 500\degree C to 201-500, an increase by a factor of 10 (Fig.~\ref{Comb_clus}(h)). In contrast, the combination method shows a decrease in the relative proportion of small clusters (11-20) from about 35\% at 500\degree C to around 22\% at 700\degree C at a dose of 0.1 dpa. Nevertheless, the predominant cluster size remains 21-50 (Fig.~\ref{Comb_clus}(i)), highlighting a significant difference between the traditional OKMC and the combination method. Clusters larger than 200 are nearly absent in the combination method, whereas they constitute 40\% of clusters in the pure OKMC case (Fig.~\ref{Comb_clus}(h)). Taking a look at the fraction of mono-vacancies, it shows almost 0\% in the pure OKMC case at 700\degree C, while this value is around 10\% in the combination method, see Fig.~\ref{Comb_clus}(i). The pure MD simulations still show a quite similar evolution as at the lower temperatures, still owing to the lack of time for significant vacancy movement.

\begin{figure}[]
\begin{center}
\subfigure{\includegraphics[width=.99\columnwidth]{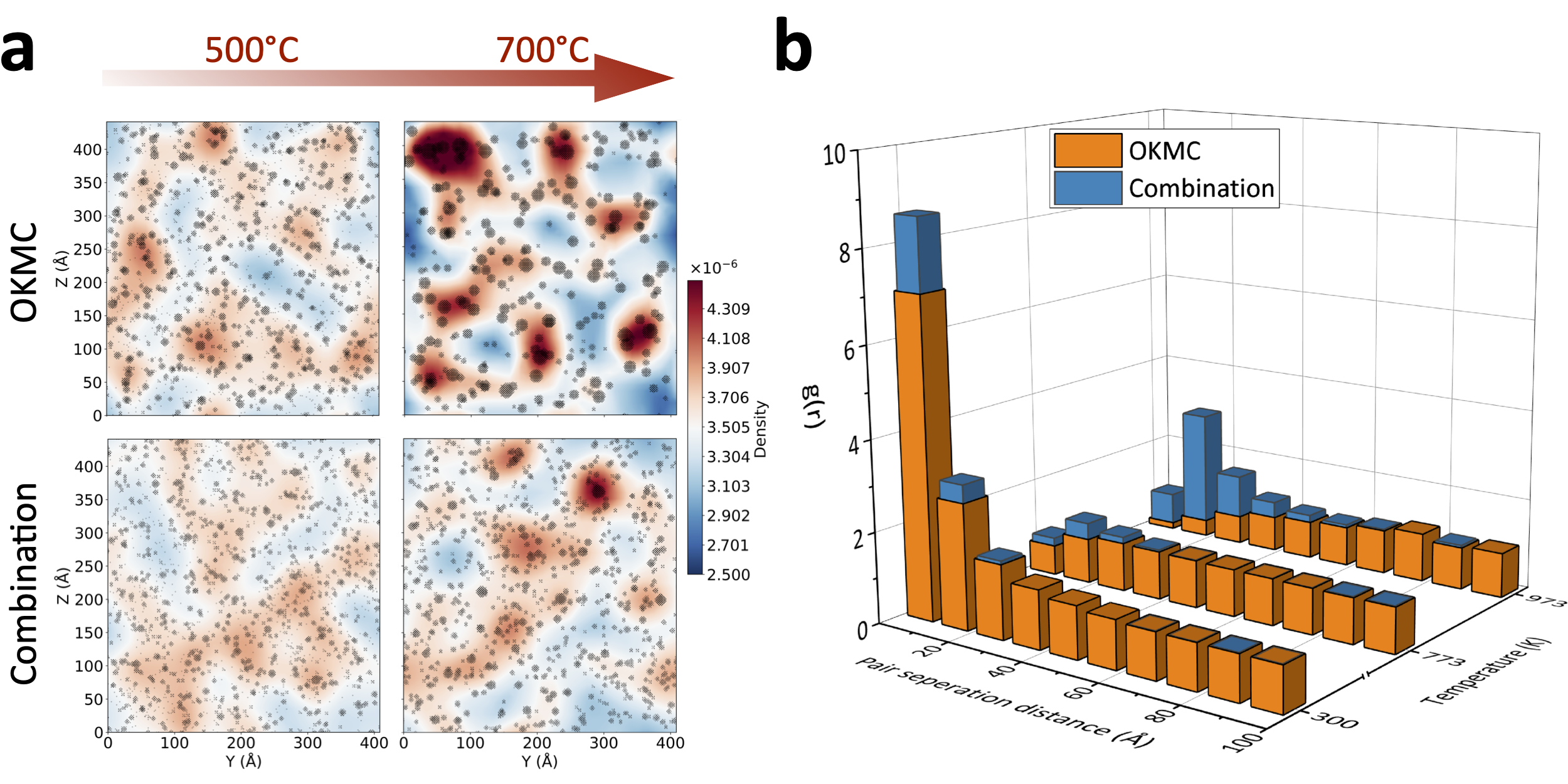}}
\end{center}
\caption{(a) Void distribution at 500\degree C and 700\degree C for OKMC and combination sets at 0.1 dpa. The black dots represent the projection of vacancy clusters aligned with the \hkl[100] direction. The color mapping shows the local density distribution of voids. (b) The spatial distribution function of voids at different temperature conditions at 0.1dpa.}
\label{gr}
\end{figure}

To quantify spatial separation, the center of mass for each void was calculated and labeled, followed by a Radial Distribution Function (RDF) analysis. Fig.~\ref{gr}(a) illustrates the void distribution in the OKMC method and in the combination method at 500\degree C and 700\degree C, at a dose of 0.1 dpa. The visualization clearly shows that the average size of voids shrinks dramatically when going from OKMC to the combination at higher temperatures. Generally, a more homogeneous distribution of voids is observed in the combination case. This difference in spatial distribution of voids is quantified by the radial distribution function, shown in Fig.~\ref{gr}(b). 

At RT, the ultra-high value of \( g(r) \) when the pair separation distance is less than 20 Å indicates a close distance between clusters (here we consider mono-vacancies also as clusters). The higher value of \( g(r) \) in the combination case at RT indicates the full cascades ``bring" the vacancies to be closer. In comparison, at 500\degree C, the difference between the OKMC and combination methods is minimal. However, the slightly higher \( g(r) \) value at short separation distances in the combination method reveals a relatively closer spatial distribution. 

At 700\degree C, the close-to-zero \( g(r) \) value at short distance in the OKMC case indicates that voids are spatially isolated from each other. In contrast, the relatively higher \( g(r) \) value at short pair distances ($<$ 40 Å) in the combination approach suggests a completely different void formation and evolution process. As evident in Fig.~\ref{gr}(a), the slight difference in density maps at 500\degree C for both cases indicates a similar void distribution. However, at 700\degree C, the extremely high local density of voids is only seen in the pure OKMC case, compared to a much more homogeneous distribution in the combination approach. 

\subsubsection{Absorbers disabled scenario}

As noted earlier, a higher vacancy concentration achieved by the OKMC results in a proportionally larger difference observed with the combination method. It is natural to speculate whether, at these low concentration levels, the combination method would still significantly impact defect formation. Comparing studies by Li et al.~\cite{li2021investigating} and Ma et al.~\cite{ma2024initial}, there is nearly a threefold difference in vacancy concentration at 0.2 dpa, due to different sink settings used. Previous sections have demonstrated that using only virtual GBs as sinks underestimates defect concentration in irradiated W at experimental relevant doses.

\begin{figure}[]
\begin{center} 
\subfigure{\includegraphics[width=.9\columnwidth]{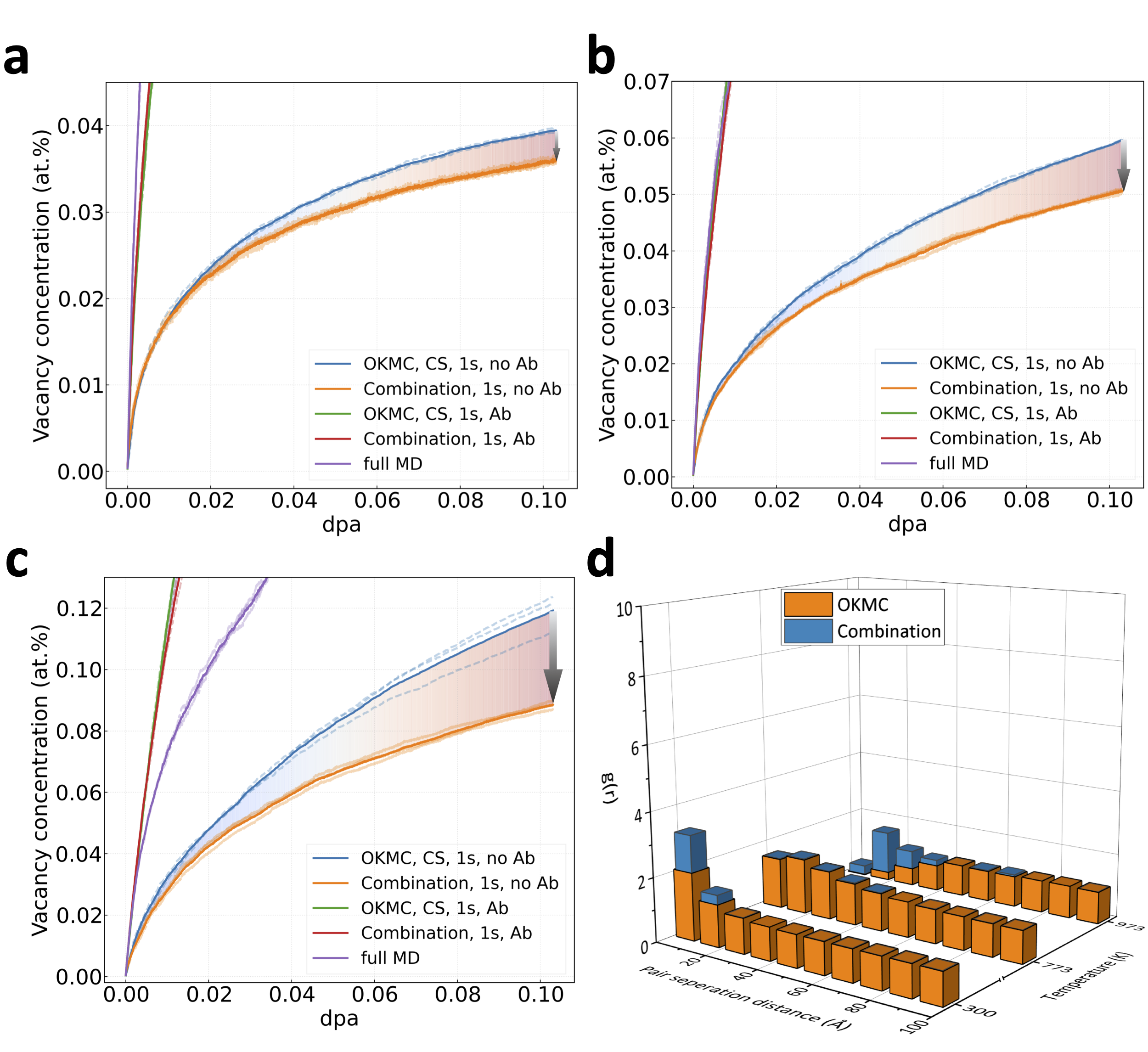}} 
\end{center}
\caption{Vacancy concentration as a function of dose at (a) RT, (b) 500\degree C and (c) 700\degree C. The curves for the absorber-enabled scenario and the full MD simulation are included for reference. (d) The spatial distribution function at different temperatures at 0.1 dpa. Dashed lines depict results from three independent runs, while the solid line shows the average value.}
\label{grnab}
\end{figure}

In this context, we performed the combination method with absorbers disabled in the OKMC part (i.e., only virtual GBs were applied as sinks). All other settings were identical, with the relaxation duration set to 1 second. Fig.~\ref{grnab} shows the evolution trend and corresponding spatial distribution of voids at the highest dose of 0.1 dpa. 

Surprisingly, even though the vacancy concentration is fully underestimated at RT, exhibiting a pseudo-saturation, the combination method still reduces the level slightly. The difference between the two approaches increases with the dose, as shown in Fig.~\ref{grnab}(a). This difference also grows with rising temperature, as depicted in Figs.~\ref{grnab}(b) and \ref{grnab}(c). A similar trend, at a much higher vacancy concentration, is observed when absorbers are enabled, shown in Fig.~\ref{Overall}. 

The spatial distribution shown in Fig.~\ref{grnab}(d) mirrors the trend in Fig.~\ref{gr}(b), except for the noticeable decrease in the absolute value of \( g(r) \) for each temperature case. This decrease is, again, due to the strong recombination between interstitials and vacancies when spherical absorbers are not applied. Interestingly, at 500\degree C, the OKMC and combination methods exhibit a very similar distribution trend. A closer inspection reveals a slight, almost indistinguishable, closer spatial distance in the combination method. The most significant difference between the two methods appears at 700\degree C, shown in Fig.~\ref{grnab}(d), indicating that the void splitting effect of full cascades dramatically changes void morphology.

\subsubsection{Comparison at different dose rates}

\begin{figure}[]
\begin{center} 
\subfigure{\includegraphics[width=.9\columnwidth]{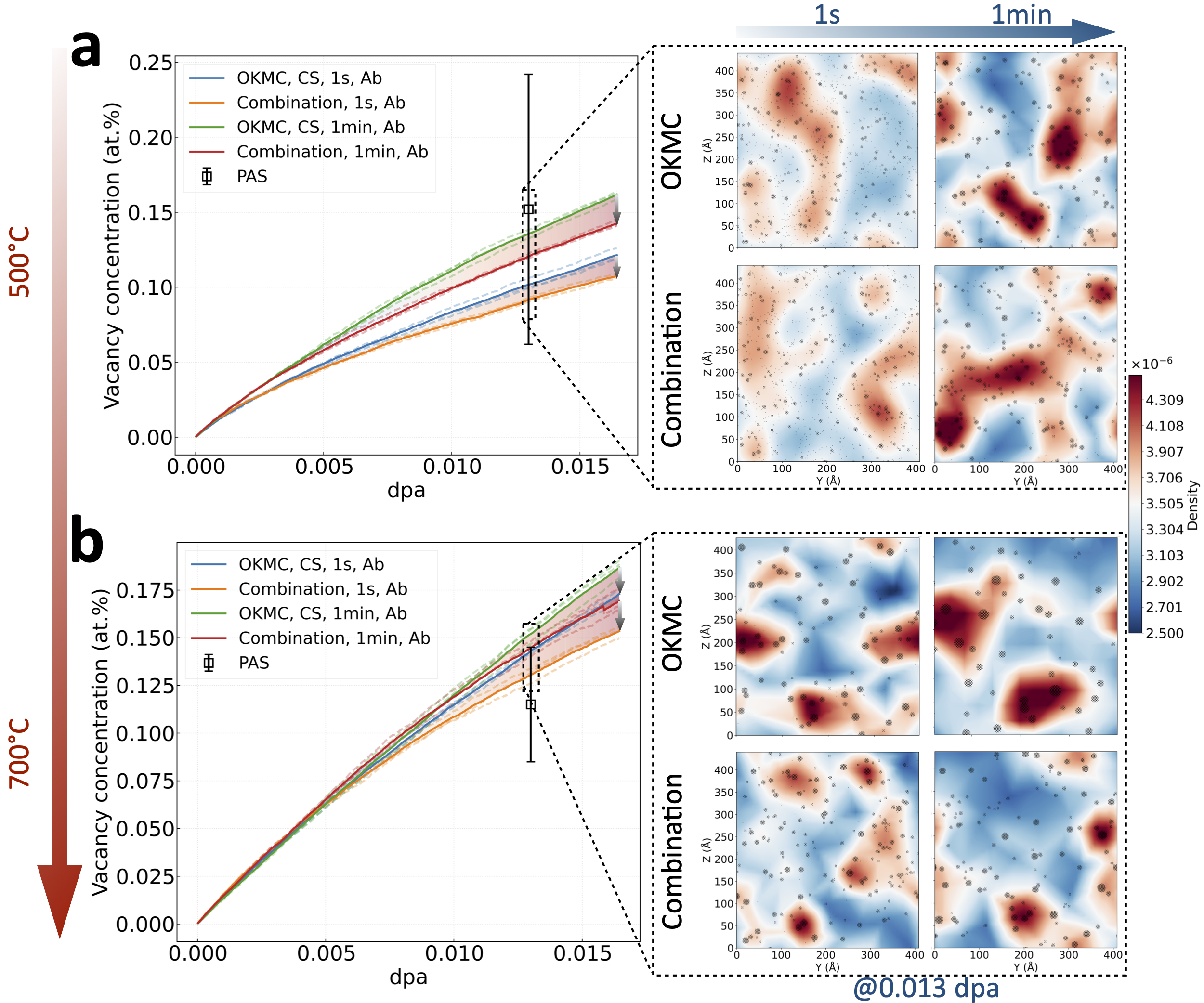}} 
\end{center}
\caption{Vacancy concentration evolution at (a) 500\degree C and (b) 700\degree C. The right panel shows the corresponding void distribution at different dose rates at 0.013 dpa. Dashed lines depict results from three independent runs, while the solid line shows the average value. Black dots represent the projection of vacancy clusters aligned with the \hkl[100] direction. The color map illustrates the local density distribution of voids. The PAS raw data points (\emph{S}-\emph{W} parameters) are obtained from Ref.~\citenum{hu2021effect}, and the vacancy concentration estimated in this study by the SA method~\cite{letter}.}
\label{1min_comb}
\end{figure}

Another point of interest is the dose rate effect in the combination method. From Fig.~\ref{VacFrac}, we observe that lower dose rates result in larger voids. We are, therefore, interested in how this affects cases when full cascades are introduced. Fig.~\ref{1min_comb} shows the vacancy concentration evolution at 500\degree C and 700\degree C. The reduction in vacancy concentration at 500\degree C slightly increases as the relaxation time is extended from 1 second to 1 minute, as shown in Fig. \ref{1min_comb}(a). Additionally, voids become larger and more local in the 1-minute case at 500\degree C. The combination method shows a denser local void distribution, with smaller voids compared to the pure OKMC case, seen in the right panel of Fig.~\ref{1min_comb}(a). On the contrary, the vacancy concentration difference between the two methods shows a decrease when the dose rate becomes lower at 700\degree C, shown in Fig.~\ref{1min_comb}(b). 

These findings reflects two competing effects: the mobility of voids at high temperatures promotes clustering behavior, while the splitting effect caused by the full cascade suppresses the size of clusters when the cascade happens on the top of them, influencing the overall evolution. The divergence between the OKMC and combination methods appears almost at the beginning of irradiation for both 1-second and 1-minute durations at 500\degree C. However, this difference becomes smaller at 700\degree C, shown in Fig.~\ref{1min_comb}(b). There is almost no difference between the OKMC and the combination methods when the dose is below 0.01 dpa. A slight difference begins to appear afterward, with an increasing trend of divergence. Coincidentally, the 1-second OKMC case and 1-minute combination case show very similar concentration levels. However, examining the void distribution (right panel of in Fig.~\ref{1min_comb}(b)), the 1-minute combination case has a higher proportion of small vacancy clusters than the 1-second OKMC case, indicating entirely different clustering behavior. A more detailed discussion of cluster size distribution at different dose rate conditions compared to the experiments will be given in the next section.

\subsubsection{Comparison with experiments}
\label{exp}

\begin{figure}[]
\begin{center}
\subfigure{\includegraphics[width=.7\columnwidth]{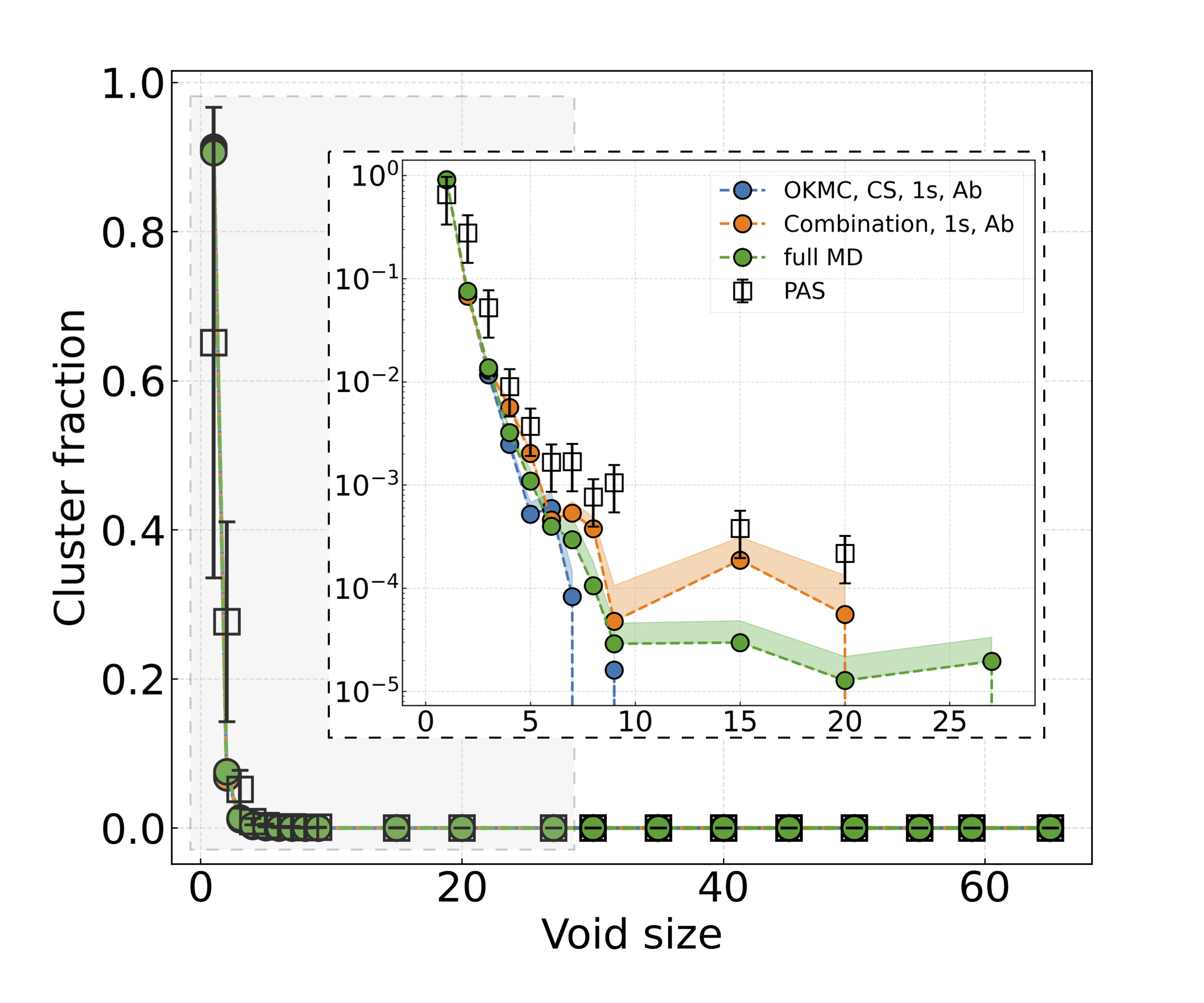}} 
\end{center}
\caption{Vacancy cluster fraction at RT, 0.05 dpa. The inset figure shows a zoomed-in view of the highlighted region. The PAS data is derived from the raw \emph{S}-\emph{W} parameters obtained from Ref.~\citenum{W_PAS_exp_new} combining with the SA algorithm.}
\label{RT_pas}
\end{figure}

To evaluate the validity of our new combination method, we compare our predictions with experimental data. Fig.~\ref{RT_pas} shows the vacancy cluster fraction at RT at 0.05 dpa. For the RT case, the surface annihilation characteristics of PAS are quite similar to those of the most damaged layer~\cite{W_PAS_exp_new}, significantly influencing the total trapping rate $k_{tot}$ and the estimation of the absolute concentration value~\cite{W_PAS_exp_new}. Therefore, we focus on the cluster fraction to quantify the void distribution in different scenarios at RT. The pure OKMC method lacks large clusters, as reflected in Fig.~\ref{Comb_clus}(b). It is evident that models incorporating the full cascade result in a larger fractions of large clusters, demonstrated by both the combination method and the full MD cases. Notably, the combination method aligns more closely with the PAS results than the full MD case. It is important to note that the cluster distribution obtained from the PAS experiment is derived from a mathematical methodology based on positron trapping model, indicating the most likely distribution to achieve the experimental \emph{S}-\emph{W} parameters, despite potential missing physical driving forces. The excellent agreement between the combination method (and the better alignment of pure MD) and the experimental results underscores the crucial role played by the full cascade.

\begin{figure}[]
\begin{center}
\subfigure[500 \degree C, 0.013 dpa]{\includegraphics[width=.49\columnwidth]{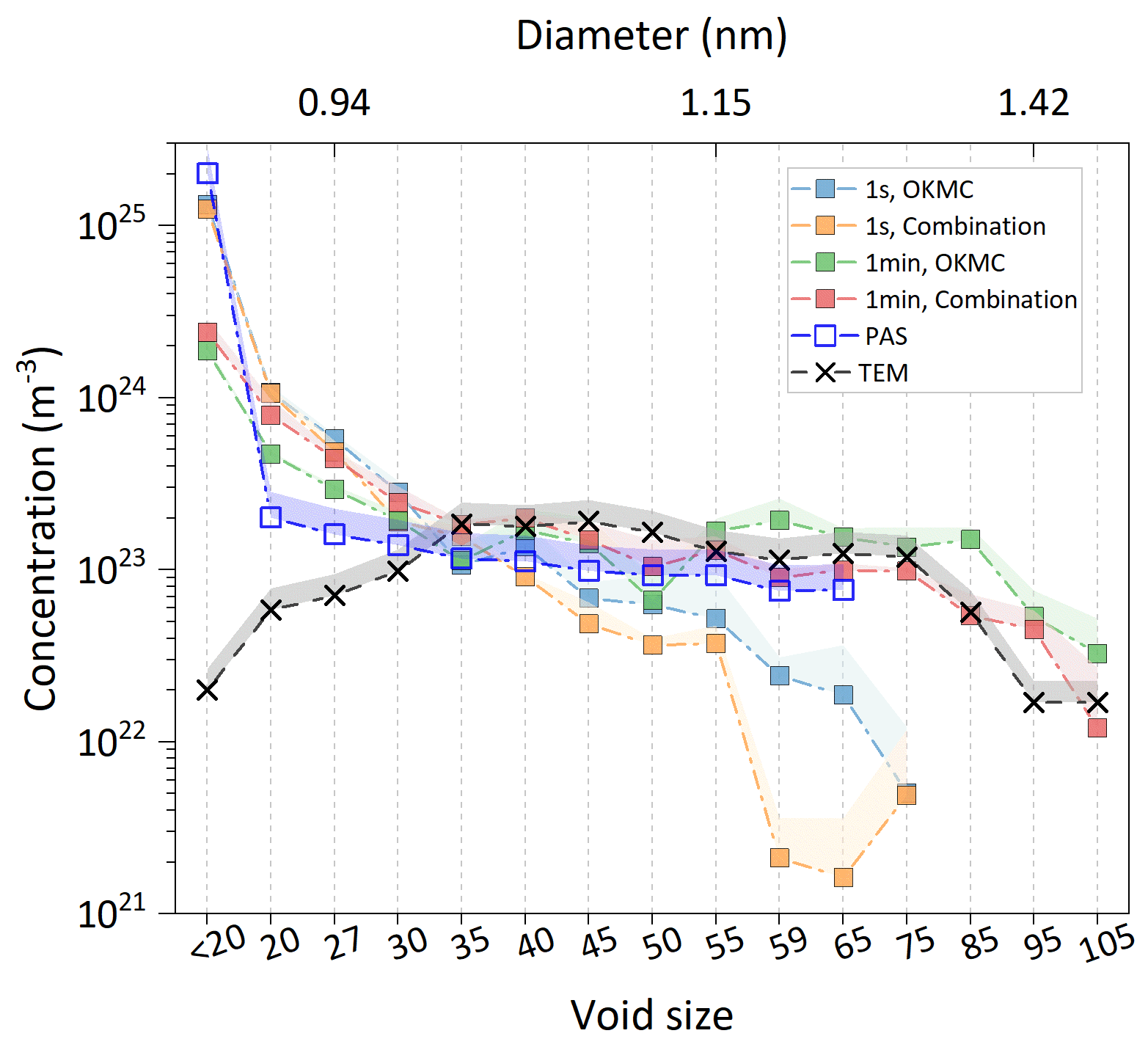}}
\subfigure[500 \degree C, 0.025 dpa]{\includegraphics[width=.49\columnwidth]{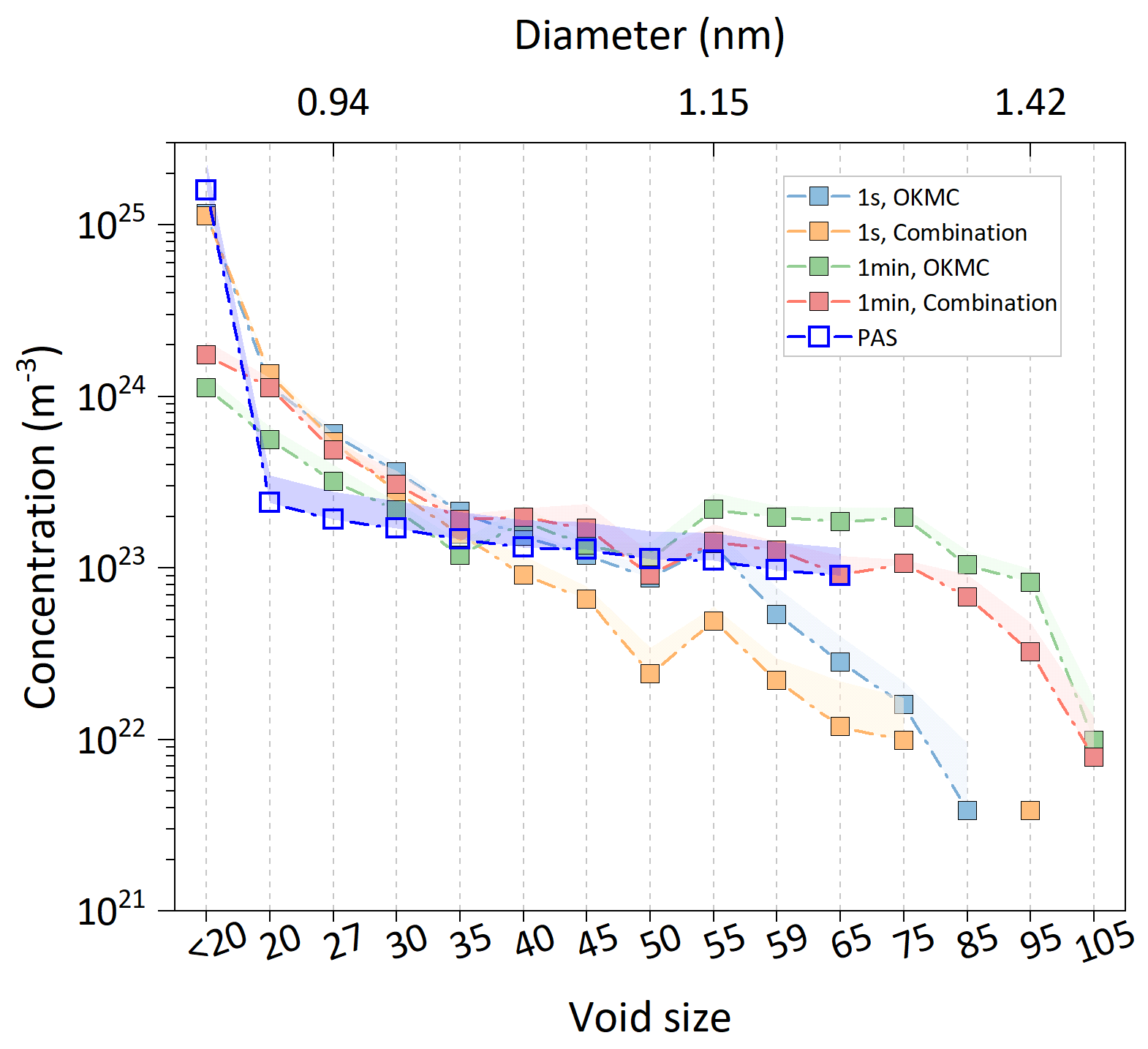}} \\
\subfigure[700 \degree C, 0.013 dpa]{\includegraphics[width=.49\columnwidth]{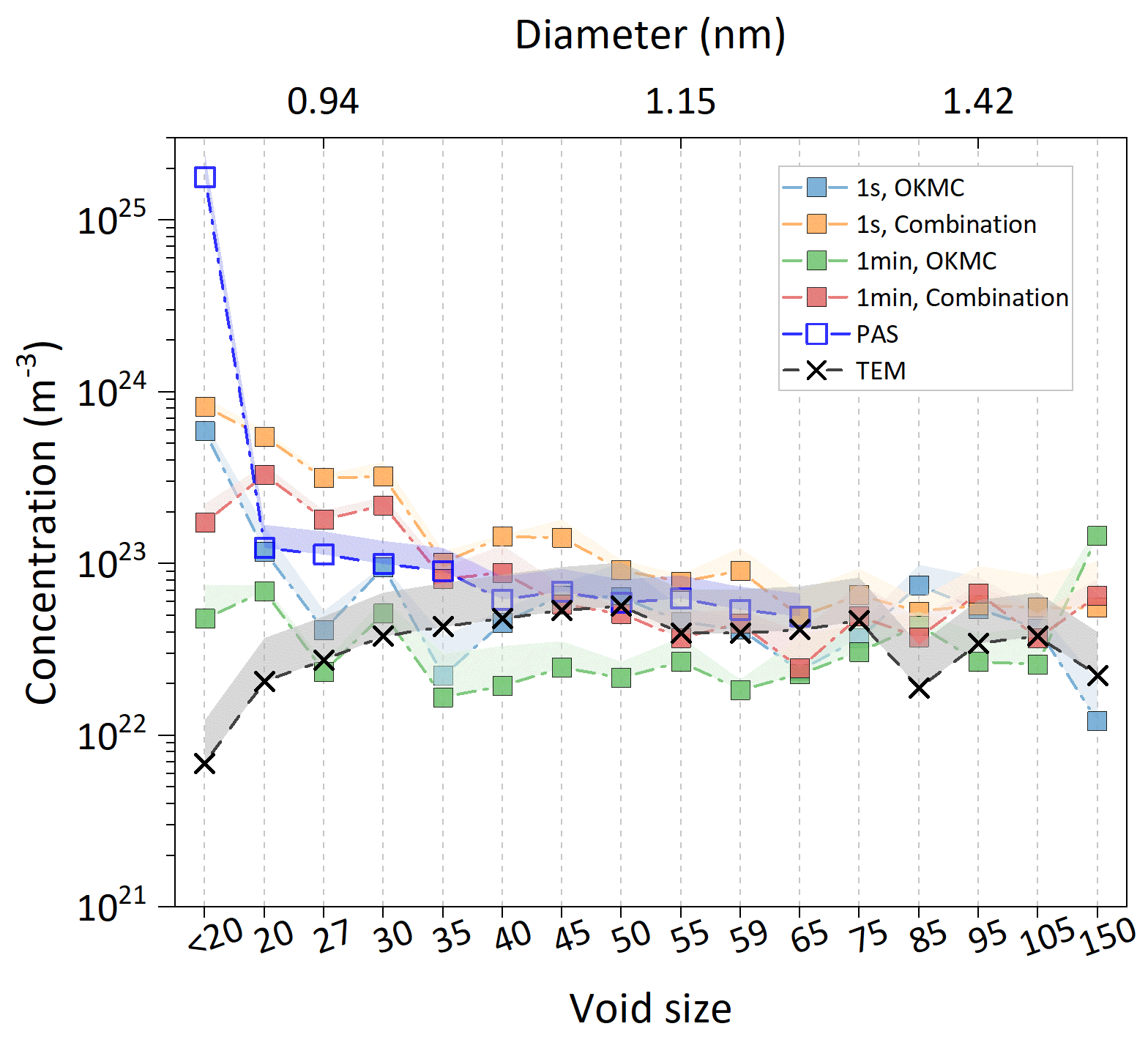}}
\end{center}
\caption{Void concentration distribution at (a) 500\degree C, 0.013 dpa, (b) 500\degree C, 0.025 dpa, and (c) 700\degree C, 0.013 dpa. All TEM and PAS data are sourced from Ref.~\citenum{hu2021effect,Huthesis} and analyzed with the SA method.}
\label{HT_pas}
\end{figure}

Based on Figs.~\ref{Overall}(c) and (d), we know that at 500\degree C, the experimental data aligns more closely with the 1-minute duration cases. A detailed discussion of the possible reasons behind this will be provided in Section~\ref{discussion}. Here, we compare the OKMC and the combination method with different dose rates to the experimental data for the higher temperature cases, shown in Fig.~\ref{HT_pas}.

First, we compare different doses at the same temperature, i.e., 500\degree C, by examining Figs.~\ref{HT_pas}(a) and (b). TEM is known to have limitations in accurately counting small clusters. However, based on the \emph{S}-\emph{W} parameters, the concentration of small clusters (with sizes smaller than 20) is represented by the 1-second cases, both using OKMC or the combination method. For larger clusters, the 1-second cases significantly underestimate the real concentration of large clusters (sizes above 55, corresponding to voids with diameters above 1.15 nm) compared to both TEM and PAS results. Upon closer inspection, for voids with sizes above 55, the 1-minute combination method aligns better with TEM results than the pure OKMC case. One reason for this is that the introduction of full cascades allows for void splitting when cascades occur on top of pre-existing voids. This effect theoretically becomes more pronounced with increasing dose, as voids progressively grow. This is evidenced by the larger concentration difference for large clusters shown at 0.025 dpa in Fig.~\ref{HT_pas}(b). Again, at this dpa level, the PAS result matches better with the 1-second cases for clusters smaller than 20, and with the 1-minute cases for larger clusters. Among these, the 1-minute combination case aligns most closely with PAS estimates in the large cluster region. 

Considering the influence of temperature when the dose is fixed at the same dpa level, i.e., 0.013 dpa, as shown in Figs.~\ref{HT_pas}(a) and (c). At 700\degree C, the combination method shows a similar distribution regardless of whether it is the 1-second or 1-minute case. However, the pure OKMC case exhibits a significant change from the 1-second to the 1-minute case: the 1-minute OKMC case dramatically underestimates the contribution of small clusters, but shows a sharp increase at cluster sizes around and above 150. In the large clusters regime, all cases, except the 1-second OKMC one, show a decent agreement with both TEM and PAS results, as shown in Fig.~\ref{HT_pas}(c). However, all simulations underestimate the concentration of small clusters (sizes less than 20) compared to the PAS estimation. The possible reasons for this will be discussed in Section~\ref{discussion}.

\subsubsection{Effect of high temperature (1077\degree C)}

\begin{figure}[]
\begin{center}
\subfigure{\includegraphics[width=.8\columnwidth]{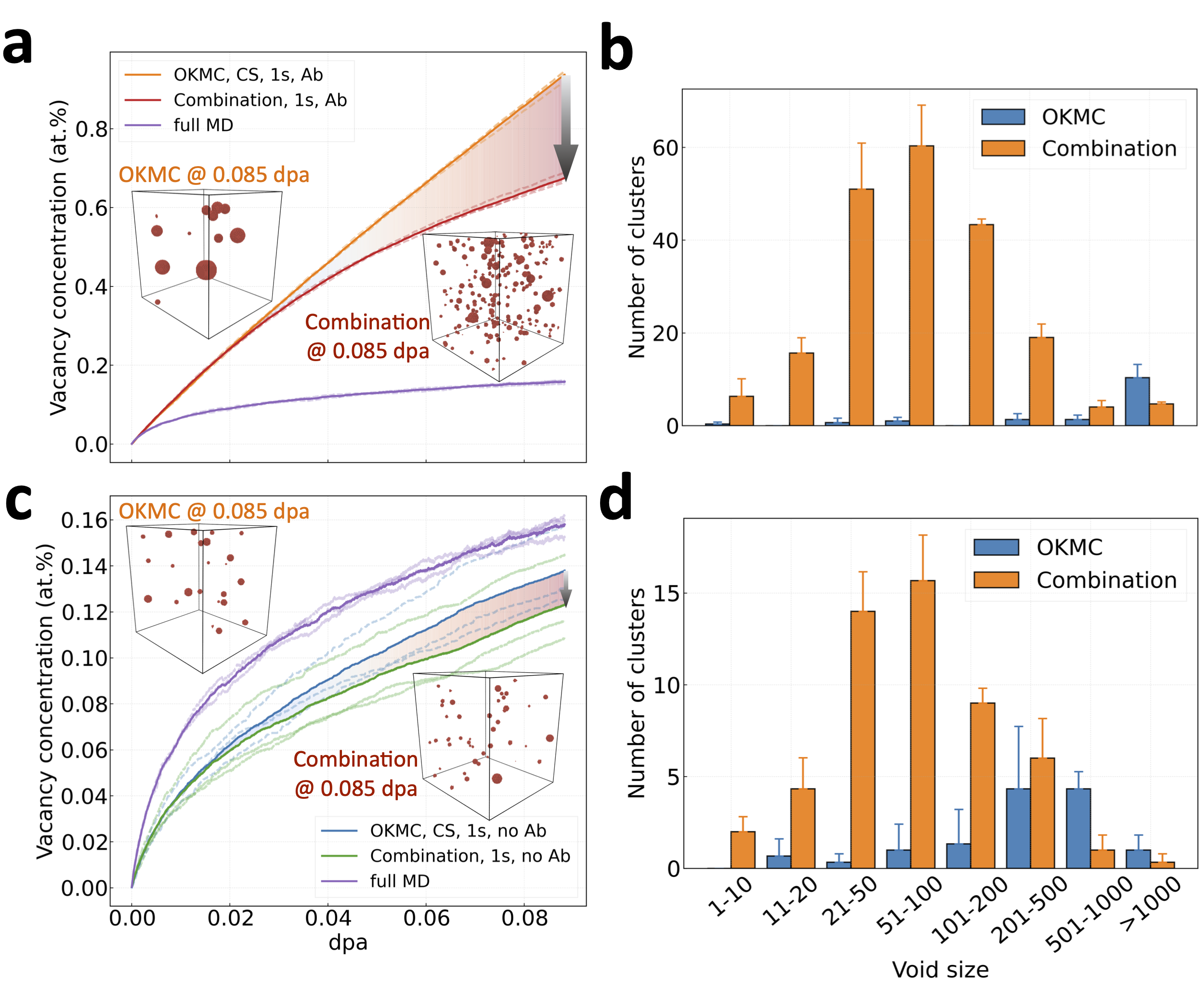}}
\end{center}
\caption{Vacancy concentration as a function of dose at 1077\degree C: (a) with and (c) without spherical absorbers in the pure OKMC, the  combination and pure MD methods. Dashed lines depict results from three independent runs, while the solid line shows the average value. The inset figures display the void distribution, shown as red atoms. (b) and (d) present the histograms of corresponding cluster distribution at the highest dose investigated in this study, 0.085 dpa at 1077\degree C.}
\label{1350}
\end{figure}

We now focus on void accumulation and evolution at a higher temperature of 1077\degree C (1350 K) by different approaches. These simulations aim to understand the performance of traditional OKMC and the combination method in such high-temperature conditions. Fig.~\ref{1350} presents the results both with and without absorbers at 1077\degree C. For reference, full MD simulations at 1077\degree C are also included.

From Fig.~\ref{1350}(a), it is evident that void accumulation starts to show an almost linear relationship with the dose in pure OKMC, when absorbers are applied. This is because, at such high temperature, vacancies are highly mobile. Therefore, if absorbers eliminate interstitials, prompt void clustering is much more likely compared to lower temperatures. Once voids grow larger, they become immobile and serve as natural ``traps" for other objects: either recombining with interstitials from subsequent irradiation or continuing to grow by trapping passing vacancies, accompanied by occasional mono-vacancy emission. As interstitials are absorbed rapidly, the clustering behavior becomes the predominate case.

As shown in the inset of Fig.~\ref{1350}(a), the pure OKMC case at 0.085 dpa shows very few but huge voids (we stopped as this dose as the voids started to grow over the size limit in OKMC, corresponding to 8 nm). Previous studies~\cite{ferroni2015high} indicate that at this temperature, visible voids under TEM conditions formed during irradiation at 500\degree C have an average diameter of around 1.5±0.1 nm after annealing at 1100\degree C. As such extreme condition it seems, albeit not a direct comparison, that the pure OKMC case overestimates the cluster sizes.

In contrast, the combination method shows a completely different void distribution, as indicated by the inset in Fig.~\ref{1350}(a). The size distribution of voids at 0.085 dpa under both approaches is quantified in Fig.~\ref{1350}(b). A Gaussian-like void size distribution appears only in the combination case, whereas the OKMC case shows much fewer clusters, with over 80\% of voids with the size larger than 1000 vacancies. Additionally, the full MD case is not shown here, as it indicates approximately 99.99 \% are clusters with size less than 10, due to the timescale limitations.

Previous sections have shown that the size distribution trend (Figs.~\ref{AborNot} and~\ref{grnab}) and evolution remain unaffected, except for the absolute concentration value, when absorbers are not applied. This holds true for the combination method at 1077\degree C, as demonstrated in Fig.~\ref{1350}(c) and (d). Fig.~\ref{1350}(c) shows much smaller clusters in the pure OKMC case without absorbers compared to Fig.~\ref{1350}(a). However, the void sizes are concentrated between 200 and 1000, corresponding to a diameter range of approximately 1.8 to 3.1 nm, as shown in Fig.~\ref{1350}(d).

The discrepancy between the cases with and without absorbers in the OKMC method indicates that the number of surviving vacancies significantly affects the results in the absence of full cascades. This finding aligns with a previous study by Bonny et al.~\cite{bonny2020trends}, which concluded that OKMC simulations at temperatures above 1000 \degree C are not feasible, as they exceed the calibrated limits of the OKMC model. In contrast, Fig.~\ref{1350}(d) demonstrates a nearly identical void distribution with a Gaussian-like shape in the combined case and when absorbers are enabled (Fig.~\ref{1350}(b)).

\section{Discussion}
\label{discussion}

Recent studies have introduced sped-up methodologies for efficiently reaching high-dose regimes in MD simulations, by replacing full cascades with FP insertions~\cite{MasonPRL,derletCRA,MasonPRM_VoidDetection}. This method has been validated at the MD timescale, if full cascades are carried out to ``anneal" out the excessive defects~\cite{Granberg2022}. However, our work demonstrates the influence of timescale, particularly when it approaches realistic dose rates, as illustrated in Fig.~\ref{HeatMap}. For example, at RT, simulations without cascade shape significantly underestimate defect concentrations (Fig.~\ref{Overall}(a)) because mobile interstitials recombine efficiently with homogeneously inserted vacancies (even though the strong sink of absorbers are used). At 700\degree C, while vacancy concentrations are similar with and without CS (Fig.~\ref{Overall}(e)), the average void size is dramatically overestimated without CS insertion (Fig.~\ref{HeatMap}(f)). It is worthy mentioning that at 700\degree C, the CS-disabled, absorber-disabled case shown in green line in Fig.~\ref{Overall}(e) matches very well with the previous work~\cite{li2021investigating}: they reported the vacancy concentration value of about 0.2 \% at 0.1 dpa.

The spherical absorbers used in this study also significantly affect defect accumulation. We observed underestimation of defect levels compared to experimental results at various temperatures (Fig.~\ref{Overall}). The similar trend also could be found from the previous studies~\cite{li2021investigating,barashev2000monte,malerba2007object,ma2024initial}. Up to 700\degree C, enabling or disabling absorbers does not affect evolution trends and clustering behavior (Figs.~\ref{AborNot} and \ref{grnab}). However, at higher temperatures, absorber settings visibly influence clustering behavior in pure OKMC simulations (Fig.~\ref{1350}). Surviving vacancies, being highly mobile, tend to cluster and become immobile traps, capturing more vacancies at this temperature. This positive feedback is more pronounced when absorbers are enabled. Despite this, absorber settings only change absolute concentration values, not evolution trends, in the combination approach (Fig.~\ref{1350}).

Our findings indicate that full cascade effects cannot simply be replaced by the capture radius definitions in traditional OKMC, especially at high temperatures. The key question we address is how to achieve a realistic void distribution comparable to experimental irradiation conditions. At RT, vacancies are immobile, and monovacancies dominate. Yet, even at this temperature, as supported by \emph{S}-\emph{W} parameter analysis~\cite{hu2021effect} combining with the SA algorithm, some larger clusters should be present to match \emph{S}-\emph{W} signals (Fig.~\ref{RT_pas}). Only MD and the combined method show large clusters at RT, absent in pure OKMC, highlighting the importance of thermal spikes for realistic defect evolution. We can observe that the full MD agrees quite well with the combination method, which indicate that pure MD can be representative for tungsten at low temperatures. As the vacancies are immobile in both full MD and the combination, the main factors at play is the cascade overlapping with the pre-existing defects, which are accounted for in both these methods.

At high temperatures, including the full cascade effect and realistic dose rates brings our void distribution models closer to actual conditions. For example, at 500\degree C, TEM and PAS results (Fig.~\ref{HT_pas}) demonstrate a clear indication of the most probable void distribution in irradiated W at a given dpa level, including both small vacancies and large voids. However, there is a discrepancy between our simulations and experimental results: for clusters smaller than size of 20, simulations with a 1-second cascade duration align better with experimental data, whereas simulations with a 1-minute cascade duration match the TEM results for larger voids.

One possible reason for this discrepancy is the constant value of \(6 \times 10^{12}/s\) used for the attempt frequency in our simulations, consistent with previous OKMC studies~\cite{niu2023influence,becquart2010microstructural,ma2024initial,li2021investigating}. In reality, the emission probability of mono-vacancies increases with the void surface area, potentially leading to an underestimation of small vacancy clusters. Additionally, despite the high purity (99.9999 wt.\%) of the experimental samples, impurities are inevitable. Previous studies~\cite{niu2023influence} show that even 5 appm of carbon impurities significantly impact void evolution by acting as natural sinks that pin dislocations, thereby indirectly reducing void size. Furthermore, the underestimation of large void sizes at comparable dose rates (1-second cascade duration) can be attributed to two factors. Experimentally, TEM observations were made after naturally cooling the sample to RT, during which voids still had time to grow. Computationally, there might be an underestimation of large cluster populations due to the size effect, as the limitations of the nano-scale boxsize increase the likelihood of cascade overlap effects.

However, the 1-minute simulations show better agreement with TEM results for large clusters when using the combination method (Fig.~\ref{HT_pas}(a)). This method also shows a higher concentration of small clusters, highlighting the crucial role of full cascades. In summary, at RT the heat spike region due to the full cascade leads to relatively larger clusters. Conversely, at high temperatures, the cascade primarily splits large voids into smaller ones. This effect become much more obvious with the temperature is increased to very high, as seen in Fig.~\ref{1350}.

Additionally, our approach could be universally applied to other materials. The material case in this study is relatively simple because interstitials are highly mobile, and recombination or absorption by GBs/absorbers occurs frequently. Despite this, the full cascade effect significantly influences defect evolution under all conditions, from RT to extremely high temperatures. We anticipate that this methodology will reveal even more unpredictable phenomena when applied to other materials. For instance, in iron, immobile C15 interstitial-type clusters~\cite{marinica2012irradiation} can significantly alter microstructural evolution, even without considering additional sink strengths like impurities or absorbers. C15 clusters themselves can serve as natural sinks, trapping and recombining with mobile vacancies. The impact of the full atomistic cascade effect combined with long timescale simulations on defect accumulation remains an unanswered question, necessitating further investigation.

\section{Conclusion}

The problem of classical MD simulations is that all conventional MD simulations are restricted to mainly nanosecond time scales, leading to orders of magnitude higher dose rates than those in experiments or reactors. To obtain experimental dose rates, larger scale models, like OKMC, are one of the options. However, since OKMC only deals with objects, it lacks some atomistic effects, such as cascades. It is known that the exact defect structures and morphologies at the atomistic level determine the macroscopic behavior of materials. Building on this and to address deficiencies of existing simulation models, we introduce a new approach: replacing the database of single PKAs, used for defect introduction in OKMC, by performing real MD-based cascades on the cells obtained from the OKMC model, to achieve reactor-relevant doses with the correct irradiation dose rates at the atomistic level. The main conclusions can be drawn as follows:
\begin{enumerate}
    \item By comparing the vacancy concentration levels to experiments, it is found that spherical absorbers could be key to obtaining realistic defect concentrations in the W case. The very mobile interstitials dramatically recombine with vacancies, leading to an underestimation of the real concentration. However, this does not affect the cluster fraction distribution under different circumstances.
    
    \item Introducing MD-based full cascades into the OKMC model significantly affects defect concentration levels, with this difference increasing with temperature. The higher the temperature, the easier it is for large clusters to form. When full cascades occur on top of pre-existing voids, they split into debris, further influencing stability -- an effect lacking in traditional OKMC, which only uses capture radius to represent cascade overlapping effects.
    
    \item Apart from the void-splitting effect, real cascades also cause a notable difference at RT irradiation, where vacancies are immobile. Absolute vacancy concentration always becomes lower with full cascades (even in cases without spherical absorbers, where only few vacancies remain). However, the combination method still shows a higher proportion of medium-sized clusters than MD simulations, indicating that realistic dose rates combining with atomistic cascades play a crucial and unique role in the defect accumulation process. The full MD and the full MD part in our combination is seen to, especially at the lowest temperature, to cluster together vacancies to voids to some extent. This shows that the full cascade will introduce two competing mechanisms, both combining monovacancies and small clusters to larger ones, as well as splitting large voids into smaller voids.
    
    \item Combining TEM-observed void distribution and mathematical method based on positron trapping model obtained distribution from PAS data, traditional OKMC tends to overestimate the proportion of large clusters compared to when full cascades are introduced. This difference becomes evident at extremely high temperatures, such as 1350K, where only the combination method results in a Gaussian-like void distribution, while the pure OKMC is not feasible anymore as they exceed the calibrated limits of the model.
\end{enumerate}

We have shown that introducing the full MD cascade to the OKMC model will affect the atomistic evolution of irradiated systems at reactor-relevant dose rates and doses. We found that having the full cascade event will dramatically affect the obtained void distribution in tungsten. The main mechanism found was that the cascade can split and destroy existing voids, which is not usually possible, if not a predefined event, in conventional OKMC simulations. The full MD cascade will take into account all the possible events that can happen during the irradiation, without the prior knowledge of them. This means that all parametrization due to cascades are automatically taken care of by the MD part. We have shown that this combination methodology will already affect the defect evolution in the quite simple material tungsten. This methodology can easily be adopted to a wide range of materials, and its effect can be much greater in materials with more complex cascade induced defect transformations. 

\section*{Acknowledgements}
This work has been carried out under the DEVHIS project, funded by the Academy of Finland (Grant number 340538). This work has partially been carried out within the framework of the EUROfusion Consortium, funded by the European Union via the Euratom Research and Training Programme (Grant Agreement No 101052200 — EUROfusion). Views and opinions expressed are however those of the author(s) only and do not necessarily reflect those of the European Union or the European Commission. Neither the European Union nor the European Commission can be held responsible for them. Computer time granted by the IT Center for Science -- CSC -- Finland is gratefully acknowledged.

\bibliographystyle{apsrev4-2}
\bibliography{mybib}

\end{document}